\documentclass[twocolumn,superscriptaddress,reprint,nofootinbib,prd]{revtex4-1}%
\usepackage{amssymb}
\usepackage{color}
\usepackage{graphicx}
\usepackage{dcolumn}
\usepackage{bm}
\usepackage[header,title,page,titletoc]{appendix}
\usepackage{amsmath}
\usepackage{bbm}
\usepackage{amsfonts}%
\usepackage[dvipdfm,bookmarks,colorlinks=false]{hyperref}
\providecommand{\U}[1]{\protect \rule{.1in}{.1in}}
\newcommand{\beq}{\begin{eqnarray}}
\newcommand{\eeq}{\end{eqnarray}}
\newcommand{\be}{\begin{equation}}
\newcommand{\ee}{\end{equation}}
\newcommand{\bw}{\begin{widetext}}
\newcommand{\ew}{\end{widetext}}
\newcommand{\ba}{\begin{array}}
\newcommand{\ea}{\end{array}}
\newcommand{\bk}{\mathbf{k}}

\newcommand{\bsigma} {{\mbox{\boldmath$\sigma$}}}
\newcommand{\bn}{\mathbf{n}}

\newcommand{\bq}{\mathbf{q}}
\newcommand{\bP}{\mathbf{P}}

\newcommand{\bfr}{\mathbf{r}}
\newcommand{\bv}{\mathbf{v}}
\newcommand{\bzero}{\mathbf{0}}
\begin{document}
\title{Generalized L\"uscher Formula in Multi-channel Baryon-Meson Scattering}

\author{Ning Li}
\affiliation{%
School of Physics, Peking University, Beijing 100871, P.~R.~China
}%

\author{Chuan Liu}%
\email[Corresponding author. Email: ]{liuchuan@pku.edu.cn}
\affiliation{%
School of Physics and Center for High Energy Physics, Peking
University, Beijing 100871, P.~R.~China
}%

\begin{abstract}
 L\"uscher's formula relates the elastic scattering phase shifts to the
 two-particle energy levels in a finite cubic box. The original formula was
 obtained for elastic scattering of two massive spinless particles in the
 center of mass frame. In this paper, we consider the case for the scattering of
 a spin $1/2$ particle with a spinless particle in multi-channel
 scattering. A generalized relation between the energy of two particle system and
 the scattering matrix elements is established. We first
 obtain this relation using quantum-mechanics in both center-of-mass frame
 and in a general moving frame. The result is then generalized to
 quantum field theory using methods outlined in Ref.~\cite{Hansen:2012tf}.
 We verify that the results obtained using both methods are equivalent up to
 terms that are exponentially suppressed in the box size.
\end{abstract}


\maketitle

\section{INTRODUCTION}

 Low-energy hadron-hadron scattering plays an important role for the
 understanding of strong interaction. However, due to its
 non-perturbative nature, it should be studied using a
 non-perturbative method like lattice Chromodynamics (lattice QCD).
 Lattice QCD can tackle the problem from first principles of QCD
 using numerical simulations. By measuring appropriate correlation
 functions, energy eigenvalues of two-particle states in a finite box
 can be obtained. L\"uscher found out a relation, now commonly known as
 L\"uscher's formula, which relates the energy of two-particle state
 in a finite box of size $L$, $E(L)$, to the elastic scattering phase
 $\delta(E(L))$ of the two particles in the
 continuum~\cite{luscher86:finiteb,luscher90:finite,luscher91:finitea,luscher91:finiteb}.
 While the former could be obtained in lattice QCD simulations, the
 latter fully characterizes the scattering property of the two
 particles and could in principle be measured in corresponding
 experiments. Thus, this relation opens up the possibility of lattice
 study of hadron-hadron scattering.

 L\"{u}scher's formalism has been utilized in a number of lattice applications, e.g.
 linear sigma model in broken phase~\cite{Gockeler:1994rx}, hadron-hadron
 scattering both with quenched approximation and unquenched configuration that
 contains dynamic
 quarks~\cite{Gupta:1993rn,Fukugita:1994ve,Aoki:1999pt,Aoki:2002in,Liu:2001ss,Hasenfratz:2004qk,Du:2004ib,Aoki:2005uf,Aoki:2002ny,Yamazaki:2004qb,Beane:2005rj}.%
 However, the original L\"{u}scher's
 method is restricted to elastic scattering of massive, spinless particles in
 the center-of-mass (COM) frame of the two particles.
 Some of these constraints restrain the applicability of the formalism to general
 hadron scattering. For example, in real simulations, one is usually restricted to
 only a few lattice
 volumes and since the energies in the box is quantized, and taking into account
 the fact that excited energy states
 are more difficult to measure numerically, one ends up with a rather poor
 energy resolution when compared with the experiments.
 To overcome this difficulty, one can of course consider
 asymmetric volumes~\cite{Li:2003jn,Feng:2004ua,Li:2007ey}, or boosting the
 system to a frame that is different from COM~\cite{Rummukainen:1995vs,XuFeng:2011,Davoudi:2011md,ZiwenFu2012,Gockeler:2012yj},
 both will enhance the energy resolution of the problem. Another possible
 generalization is to use the so-called twisted boundary
 conditions advocated in
 Refs.~\cite{Bedaque:2004ax,Bedaque:2004kc,Sachrajda:2004mi,deDivitiis:2004kq}.

 Generalizations to particles with
 spin~\cite{Beane:2006mx,Beane:2003da,Meng:2003gm} is also possible. For
 example, in Ref.~\cite{Bernard:2008ax,Ishizuka:2009bx}, L\"{u}scher's formula
 has been extended to elastic scattering of baryons.

 It is also important to extend L\"uscher's formula to the case of
 inelastic scattering which is commonly encountered in hadronic physics.
 Attempts have been made over the years, see
 Refs.~\cite{He:2005ey,Liu:2005kr,Lage:2009zv,Ishii:2011tq,Aoki:2011gt,Hansen:2012tf,Doring:2011vk,Doring:2012eu}.

 In this paper, we would like to synthesize the above mentioned generalizations
 by trying to seek a formula (or a formalism) that is applicable for
 multi-channel scattering of particles with spin in a possibly moving frame (MF).
 For this purpose, we consider two-particle to two-particle
 scattering processes in which one particle has spin $1/2$ (we will henceforth call it a ``baryon'')
 while the other particle remains spinless (we will call it a ``meson'').
 The scattering process we study could be a multi-channel
 scattering beyond a certain threshold. But in each channel, the
 two-particle states (initial or final) always has the feature that
 one particle has spin $1/2$ while the other is spinless.
 We will comment on possible extensions to this restriction in
 Sec.~\ref{sec:conclude}.
 As it turns out, the basic formulae obtained are the same in
 non-relativistic quantum mechanics and in massive quantum field theories, we
 will start our discussion in the former case. Generalization to
 quantum field theory can be achieved by using methods outlined in
 Refs.~\cite{Ishizuka:2009bx,Hansen:2012tf} which is detailed
 in Sec.~\ref{sec:QFT}.

 The organization of the paper is as follows: In Sec.~\ref{sec:QM},
 we start out by discussing a quantum-mechanical model. L\"uscher's  formulae are
 obtained first in the case of COM frame in
 subsection~\ref{subsec:COM} and then generalize to moving frames
 in subsection~\ref{subsec:MF}. In Sec.~\ref{sec:QFT}, we
 generalize the results obtained in Sec.~\ref{sec:QM} to quantum
 field theory. This is achieved first in the single channel situation
 and then to the two-channel scenario. We then compare the results obtained
 in quantum field theory with those from quantum mechanics in Sec.~\ref{sec:QM}.
 It is shown that they are in fact equivalent up to terms that are
 exponentially suppressed in the large volume limit.
 In Sec.~\ref{sec:conclude}, before we conclude, we will also discuss possible
 applications of our formulae in real lattice simulations and comment on
 some possible extensions in the future.

 Some calculation details are summarized in the
 appendices. To be more specific, single channel scattering
 for particles with spin in infinite volume are reviewed in
 appendix~\ref{appendix:spin_scattering}. For reference, single channel L\"uscher's formulae
 are also provided in this appendix. Calculations of
 loop summation/integration in the case of quantum field theory
 are summarized in appendix~\ref{appendix:loops}.

 \section{L\"{u}scher's formula for baryon-meson scattering in non-relativistic quantum mechanics}
 \label{sec:QM}

 \subsection{L\"{u}scher's formula in COM frame}
 \label{subsec:COM}

 \subsubsection{Two-channel scattering in the continuum}
 In this section, using non-relativistic quantum mechanics in COM frame, we will
 briefly discuss two-channel potential scattering in the continuum
 for the case of two stable particles: one with spin $0$ (a meson)
 and the other with spin $\frac{1}{2}$ (a baryon).
 We will follow the discussion in Ref.~\cite{He:2005ey}. The potential
 $V(r)$ between the two particles is assumed
 to have finite range, i.e., $V(r)=0$ with $r=|\mathbf{r}|>R$
 for some positive $R$, but the potential itself could in principle
 be spin-dependent so that the spin of the fermion might change
 during the scattering process. We assume that there exists a threshold $E_T>0$
 and the energy of the two-particle system becomes
 \beq
 E&=&{\bk^2_1\over 2\mu_1}=E_T+{\bk^2_2\over 2\mu_2}\;,
 \eeq
 where $\mu_1$ and $\mu_2$ are the reduced mass of the two-particle
 system below and above the threshold, respectively.
 In the COM frame, one only has to denote the momentum
 of one of the two particles. For definiteness,
 we denote the momentum of the fermion
 as $\bk_1$ and $\bk_2$ in the first and second channel, respectively.
 The magnitude of them are $k_1=|\bk_1|$ and $k_2=|\bk_2|$.
 Obviously, for energies below the threshold
 $k_2$ will become pure imaginary.

 The wave function of two-particle system, after factoring out
 the trivial COM coordinates, has a
 two-component form in the case of
 two-channel scattering:~\cite{He:2005ey}
 \beq
 \label{eq:eins}
 \psi(\mathbf{r})&=&\left(
 \ba
[c]{c}%
\psi_{1}(\mathbf{r})\\
\psi_{2}(\mathbf{r})
 \ea
 \right)\;.
 \eeq
 Note that due to spin degrees of freedom, each component is still
 a two-component spinor.
 At large $r$ where the potential $V(r)$ vanishes, the wave
 functions of the scattering states can be chosen to have the following forms:
 \beq
 \psi_{1;s}(\mathbf{r})\overset{r\rightarrow \infty}{\longrightarrow}\left(
 \ba
 {c}
 \chi^{\frac{1}{2}}_se^{i\mathbf{k_{1}\cdot{r}}}
 +\sum_{s^{\prime}} \chi^{\frac{1}{2}}_{s^{\prime}}M^{(NR)}_{11;s^{\prime}s}
\frac{e^{ik_1r}}{r}\\
 \sqrt{\frac{\mu_2}{\mu_1}}\sum_{s^{\prime}}\chi^{\frac{1}{2}}_{s^{\prime}}M^{(NR)}_{21;s^{\prime}s}
 \frac{{e^{ik_2r}}}{r}
 \ea
 \right)\;,
 \eeq
 \newline
 \beq
 \psi_{2;s}(\mathbf{r})\overset{r\rightarrow \infty}{\longrightarrow}\left(
 \ba
 {c}
 \sqrt{\frac{\mu_1}{\mu_2}}\sum_{s^{\prime}}\chi^{\frac{1}{2}}_{s^{\prime}}M^{(NR)}_{12;s^{\prime}s}
 \frac{{e^{ik_{1}r}}}{r}\\
 \chi_s^{\frac{1}{2}}e^{i\mathbf{k_{2}\cdot{r}}}
 +\sum_{s^{\prime}} \chi^{\frac{1}{2}}_{s^{\prime}}M^{(NR)}_{22;s^{\prime}s}
\frac{e^{ik_{2}r}}{r}
 \ea \right)\;.
 \eeq
 \newline
 In the above expressions, $\chi^{\frac{1}{2}}_s$ is an eigenstate of
 spin angular momentum $s_{z}$ of the baryon with eigenvalue
 $s=-1/2,1/2$.  The scattering amplitudes
 like $M^{(NR)}_{ij;s's}$ depends only on the corresponding angles
 $\hat{\bk}_1\cdot\hat{\bfr}$ and $\hat{\bk}_2\cdot\hat{\bfr}$.
 In order not to confuse with the matrix $M$ introduce in quantum field theory in Sec.~\ref{sec:QFT},
 we have added a superscript (NR) to stand for the case of non-relativistic quantum
 mechanics.
 The notation in this paper is as follows. Subscripts like $i$ and $j$ refer to the channel,
 and take the values 1 or 2.
 It is seen that, in the remote past, $\psi_{1;s}(\mathbf{r})$ becomes a
 pure incident plane wave in the first channel
 with definite linear momentum $\mathbf{k_{1}}$ and definite spin
 $s$. Similarly, in the remote past,
 $\psi_{2;s}(\mathbf{r})$ represents an incident plane wave
 in the second channel with definite momentum $\bk_2$ and definite spin $s$.
 It is also clear that the above wave functions, $\psi_{i;s}(\bfr)$ with
 $s=-1/2,1/2$ are linearly
 independent and in fact they are also complete in the sense that
 any eigenfunction of the Hamiltonian must be a linear superposition
 of them.

 As is said, the scattering amplitudes $M^{(NR)}_{ij;s's}$ for
 $s',s=1/2,-1/2$ depend only on
 the corresponding angles and can be expanded into spherical harmonics,
 see for example Ref.~\cite{Newton}.
 For simplicity, we have chosen the $z$-axis to coincide with the
 incident momentum, $\bk_1$ and $\bk_2$ for $\psi_{1;s}(\mathbf{r})$
 and $\psi_{2;s}(\mathbf{r})$, respectively.
 The scattering amplitudes then takes the following form:
 \bw
 \beq
 M^{(NR)}_{11;s^{\prime}s}(\mathbf{\hat{k}_{1}\cdot{\hat{r}}})&=&\frac{1}{2ik_{1}}\sum_{l=0}^\infty\sum_{J=l-\frac{1}{2}}^{l+\frac{1}{2}}
 \sqrt{4\pi(2l+1)}
 (S_{11}^{Jl}-1)S^{l\frac{1}{2}}_{JM;ms^{\prime}}S^{l\frac{1}{2}}_{JM;0s}Y_{lm}(\mathbf{\hat{r}})
 \;,
 \\
 M^{(NR)}_{12;s^{\prime}s}(\mathbf{\hat{k}_{2}\cdot{\hat{r}}})
 &=&\frac{1}{2i\sqrt
 {k_{1}k_{2}}}\sum_{l=0}^\infty\sum_{J=l-\frac{1}{2}}^{l+\frac{1}{2}}\sqrt{4\pi(2l+1)}
 {S^{Jl}_{12}}S^{l\frac{1}{2}}_{JM;ms^{\prime}}S^{l\frac{1}{2}}_{JM;0s}Y_{lm}(\mathbf{\hat{r}})
 \;.
 \eeq
 \ew
 In the above expressions,  $S_{JM;ms^{\prime}}^{l\frac{1}{2}}=\langle{JM}|lm;\frac{1}{2}s^{\prime}\rangle$
 and $S_{JM;0s}^{l\frac{1}{2}}=\langle{JM}|l0;\frac{1}{2}s\rangle$
 are the Clebsch-Gordan (CG) coefficients.

 The quantities $S^{Jl}_{ij}$ are the $S$-matrix elements
 where $J$ is the quantum number of the total angular momentum.
 Since we are dealing with the scattering of a
 spin $1/2$ particle, $J$ can take two possible values
 for a given $l$, namely $J=l\pm 1/2$.
 To simplify the notation we will also denote the corresponding
 $S$-matrix elements by
 \be
 S^{J=l\pm 1/2,l}_{ij}\equiv S^{l\pm}_{ij}\;.
 \ee
 Needless to say, one also has similar expressions for $M^{(NR)}_{21;s^{\prime}s}$ and
 $M^{(NR)}_{22;s^{\prime}s}$. In the following,
 we will also need the so-called spin spherical harmonics defined as
 \be
 \label{eq:spin_spherical_harmonics_def}
 Y_{JM}^{l\frac{1}{2}}(\mathbf{\hat{r}})
 =\sum_{ms}Y_{lm}(\mathbf{\hat{r}})\chi_{s}^{\frac{1}{2}}S_{JM;ms}^{l\frac{1}{2}}
 \;,
 \ee
 which is an eigenfunction of the total angular momentum of the
 system. The above expressions are just direct generalizations to the
 spin-dependent single-channel scattering. For convenience, some relevant
 formulae are collected in appendix~\ref{appendix:spin_scattering}.

 With the spin spherical harmonics defined in Eq.~(\ref{eq:spin_spherical_harmonics_def}),
 we can expand the wave function in the following form:
 \be
 \psi_{i;s}(\mathbf{r})=\sum_{JMl}
 \sqrt{4\pi(2l+1)}S_{JM;0s}^{l\frac{1}{2}}W_{i;Jl}(r)
 Y_{JM}^{l\frac{1}{2}}(\mathbf{\hat{r}})\;,
 \ee
 where the radial wave functions of Schr\"{o}dinger equation
 are denoted by
 $W_{i;Jl}(r)$.
 In the large $r$ region, they have the following asymptotic forms
 \beq
 W_{1;Jl}(r)&=&\left(
 \ba
 [c]{c}
 \frac{1}{2ik_1r}[S^{Jl}_{11}e^{ik_1r}+(-1)^{l+1}e^{-ik_1r}]\\
 \frac{1}{2ir\sqrt{k_1k_2}}\sqrt{\frac{\mu_2}{\mu_1}}S^{Jl}_{21}e^{ik_2r}
 \ea
 \right)\;,\nonumber\\
 \eeq
 \beq
 W_{2;Jl}(r)&=&\left(
 \ba
 [c]{c}
 \frac{1}{2ir\sqrt{k_1k_2}}\sqrt{\frac{\mu_{1}}{\mu_{2}}}S^{Jl}_{12}e^{ik_1r}\\
 \frac{1}{2ik_2r}[S^{Jl}_{22}e^{ik_2r}+(-1)^{l+1}e^{-ik_2r}]
 \ea
 \right)\;.\nonumber\\
 \eeq
 It is obvious that two radial wave functions $W_{1;Jl}(r)$ and
 $W_{2;Jl}(r)$ are linearly independent. Since the radial
 Schr\"{o}dinger equation has two linearly independent
 solutions which are regular at the origin,
 denoted by $u_{i;Jl}(r)$, these radial wave
 functions can be expressed as linear superpositions of two radial wave
 functions $W_{i;Jl}(r)$~\cite{He:2005ey}.

 \subsubsection{Two-channel scattering in a cubic box}

 Now we put the two-particle system into a cubic box of size $L$
 and impose the periodic boundary
 condition. The potential then becomes
 $V_{L}(\mathbf{r})=\sum \nolimits_{\mathbf{n}}V(\mathbf{r}+\mathbf{n}L)$,
 $\bn\in \mathbbm{Z}^3$. We divide the whole space into
 two regions: the inner region and the outer region.
 In the inner region, every point satisfy the
 condition: $|\mathbf{r}+\mathbf{n}L|<R$, for some $\mathbf{n}\in
 \mathbbm{Z}^3$ while
 in the outer region, $\Omega=\{ \mathbf{r}|:|\mathbf{r}+\mathbf{n}L|>R$,
 $\mathbf{n}\in \mathbbm{Z}^3\}$.
 In the inner region, the solution to Schr\"{o}dinger equation of
 the system is
 \be
 \psi(\mathbf{r})=\sum_{l=0}^\infty\sum_{J=l-\frac{1}{2}}^{l+\frac{1}{2}}\sum_{M=-J}^{J}
 \left[\sum_{i=1}^{2}b_{i;JMl}u_{i;Jl}
 (r)\right]
 Y_{JM}^{l\frac{1}{2}}(\mathbf{\hat{r}})\;.
 \ee
 Since $u_{i;Jl}(r)$ can be expressed as linear
 superpositions of two radial wave functions $W_{i;Jl}(r)$, in
 the outer region the wave function is given by
 \be
 \psi(\mathbf{r})=\sum_{l=0}^\infty\sum_{J=l-\frac{1}{2}}^{l+\frac{1}{2}}
 \sum_{M=-J}^{J}
 \left[\sum_{i=1}^{2}F_{i;JMl}W_{i;Jl}(r)\right]
 Y_{JM}^{l\frac{1}{2}}(\mathbf{\hat{r}})\;.
 \label{wavefunction7}
 \ee
 with $F_{i;JMl}$ being some non-trivial constants.
 On the other hand, in the outer region
 $\Omega$, the wave function $\psi(\mathbf{r})$ is also a linear
 superposition of the so-called singular periodic solutions~\cite{luscher91:finitea} to
 the Helmholtz equation, $G_{i;JMl}(\mathbf{r};k_{i}^{2})$, also known
 as the Green's functions. Thus we also have
 \beq
 \psi(\mathbf{r})&=&\left(
 \ba
 [c]{c}%
 \overset{\infty}{\underset{l=0}{\sum}}\overset{l+\frac{1}{2}}
 {\underset{J=l-\frac{1}{2}}{\sum}}\overset{J}{\underset{M=-J}{\sum}}
 \nu_{1;JMl}G_{1;JMl}(\mathbf{r};k_{1}^{2})\\
 \overset{\infty}{\underset{l=0}{\sum}}\overset{l+\frac{1}{2}}{\underset{J=l-\frac{1}{2}}{\sum}}
 \overset{J}{\underset{M=-J}{\sum}}
 \nu_{2;JMl}G_{2;JMl}(\mathbf{r};k_{2}^{2})
 \ea
 \right)\;. \nonumber\\
 \label{wavefunction8}
 \eeq
 From Ref.~\cite{luscher91:finitea,Ishizuka:2009bx}, the Green's
 function for particles with spin takes the following form:
 \bw
 \be
 G_{i;JMl}(\textbf{r},k_i^2)=
 \frac{(-1)^lk^{l+1}_i}{4\pi}\left(Y_{JM}^{l\frac{1}{2}}(\mathbf{{\hat{r}}})n_l(k_ir)
 +\sum_{l^{\prime}=0}^\infty\sum_{J^{\prime}=l^{\prime}-\frac{1}{2}}^{l^{\prime}+\frac{1}{2}}\sum_{M^{\prime}=-J^{\prime}}^{J^{\prime}}
  \mathcal{M}_{i;JMl;J^{\prime}M^{\prime}l^{\prime}}(k^2_i)
 Y_{J^{\prime}M^{\prime}}^{l^{\prime}\frac{1}{2}}
 (\mathbf{{\hat{r}}})j_{l^{\prime}}(k_ir)
\right),
 \ee
 \ew
 where the explicit form of $\mathcal{M}_{i;JMl;J^{\prime}M^{\prime}l^{\prime}}(k^2_i)$
 can be found in appendix~\ref{appendix:spin_scattering}. For
 $\mathcal{M}_{i;JMl;J^{\prime}M^{\prime}l^{\prime}}(k^2_i)$,
 one has to substitute $k^2=k^2_i$ into
 Eq.~(\ref{eq:M_JLM}) to get
 $\mathcal{M}_{i;JMl;J^{\prime}M^{\prime}l^{\prime}}(k^2_i)$.

 To simplify the notation, we define $\tilde{F}_{1;JMl}=\sqrt{\mu_2}F_{1;JMl}$,
 $\tilde{F}_{2;JMl}=\sqrt{\mu_1}F_{2;JMl}$,
 $\tilde{\nu}_{1;JMl}=\sqrt{\mu_2}\nu_{1;JMl}$ and $\tilde{\nu}_{2;JMl}=\sqrt{\mu_1}\nu_{2;JMl}$.
 Then the equivalence of  Eqs.~(\ref{wavefunction7}),(\ref{wavefunction8}) leads to
 four linear equations involving $\tilde{F}_{i;JMl}$'s and $\tilde{\nu}_{i;JMl}$'s,
 \bw
 \be
 \label{linear_equation4}
 \left\{ \begin{aligned}
 \tilde{F}_{1;JMl}(S^{Jl}_{11}+1)+\tilde{F}_{2;JMl}\sqrt{\frac{k_1}{k_2}}S^{Jl}_{12}
 &=\sum_{J^{\prime}M^{\prime}l^{\prime}}\mathcal{M}_{1;JMl;J^{\prime}M^{\prime}l^{\prime}}\frac{(-1)^{l^{\prime}}k_1^{l^{\prime}+1}}{4\pi}
 \tilde{\nu}_{1;J^{\prime}M^{\prime}l^{\prime}}
 \\
 -i\tilde{F}_{1;JMl}(S^{Jl}_{11}-1)-i\tilde{F}_{2;JMl}\sqrt{\frac{k_1}{k_2}}S^{Jl}_{12}
 &=\frac{(-1)^{l}k_1^{l+1}}{4\pi}\tilde{\nu}_{1;JMl}\\
 \tilde{F}_{2;JMl}(S^{Jl}_{22}+1)+\tilde{F}_{1;JMl}\sqrt{\frac{k_2}{k_1}}S^{Jl}_{21}
 &=\sum_{J^{\prime}M^{\prime}l^{\prime}}\mathcal{M}_{2;JMl;J^{\prime}M^{\prime}l^{\prime}}\frac{(-1)^{l^{\prime}}
 k_2^{l^{\prime}+1}}{4\pi}
 \tilde{\nu}_{2;J^{\prime}M^{\prime}l^{\prime}}
\\
 -i\tilde{F}_{2;JMl}(S^{Jl}_{22}-1)-i\tilde{F}_{1;JMl}\sqrt{\frac{k_2}{k_1}}S^{Jl}_{21}
 &=\frac{(-1)^{l}k_2^{l+1}}{4\pi}\tilde{\nu}_{2;JMl}
 \end{aligned} \right.\;.
 \ee
 \ew
 One can eliminate the coefficients $\tilde{\nu}_{1;JMl}$  and
 $\tilde{\nu}_{2;JMl}$ easily, leaving behind a set of linear
 equations for $\tilde{F}_{1;JMl}$ and $\tilde{F}_{2;JMl}$. In order to have non-trivial
 solutions for them, the determinant of the corresponding matrix must vanish.
 Let us now define
 \be
 \mathcal{M}^{\pm}_{i;JMl;J^{\prime}M^{\prime}l^{\prime}}(k^2_i)=\mathcal{M}_{i;JMl;J^{\prime}M^{\prime}l^{\prime}}(k^2_i)
 \pm i\delta_{JJ^{\prime}}\delta_{MM^{\prime}}\delta_{ll^{\prime}}\;.
 \ee
 Using matrix $ \mathcal{M}^{\pm}_{i;JMl;J^{\prime}M^{\prime}l^{\prime}}$, we can write the secular equation for
 Eq.~(\ref{linear_equation4}) as
 \bw
 \be
 \left|
 \ba
 {cc}
 \mathcal{M}^{+}_{1;JMl;J^{\prime}M^{\prime}l^{\prime}}
 -S^{Jl}_{11}\mathcal{M}^{-}_{1;JMl;J^{\prime}M^{\prime}l^{\prime}}
 &   \sqrt{\frac{k_2}{k_1}}S^{Jl}_{21}\mathcal{M}^{-}_{2;JMl;J^{\prime}M^{\prime}l^{\prime}}\\
 \sqrt{\frac{k_1}{k_2}}S^{Jl}_{12}\mathcal{M}^{-}_{1;JMl;J^{\prime}M^{\prime}l^{\prime}}
 &
\mathcal{M}^{+}_{2;JMl;J^{\prime}M^{\prime}l^{\prime}}
 -S^{Jl}_{22}\mathcal{M}^{-}_{2;JMl;J^{\prime}M^{\prime}l^{\prime}}
 \ea
 \right|
 =0\;.
 \label{luscher_formula28}
 \ee
 \ew
 We will also use the more compact matrix notation
 $\mathcal{M}^{\pm}_{1,2}=\mathcal{M}_{1,2}\pm i$.
 Assuming that the matrix
 $\mathcal{M}^{-}_i$ is non-singular,
 we define a unitary matrice $U_i$ as
 \beq
 U_{i;JMl;J^{\prime}M^{\prime}l^{\prime}} &=&
 \left[\mathcal{M}_i^{+}\left(\mathcal{M}_i^{-}\right)^{-1}\right]_{JMl;J^{\prime}M^{\prime}l^{\prime}}
 \nonumber \\
 &=&\left(\frac{\mathcal{M}_i+i}{\mathcal{M}_i-i}\right)_{JMl;J^{\prime}M^{\prime}l^{\prime}}
 \;.
 \eeq
 With this matrix, the generalized L\"{u}scher's formula for two-channel
 baryon-meson scattering may be written in an equivalent form
 \bw
 \be
 \left|
 \ba
 {cc}
 U_{1;JMl;J^{\prime}M^{\prime}l^{\prime}}
 -S^{Jl}_{11}\delta_{JJ^{\prime}}\delta_{MM^{\prime}}\delta_{ll^{\prime}}
 &   \sqrt{\frac{k_2}{k_1}}S^{Jl}_{21}\delta_{JJ^{\prime}}\delta_{MM^{\prime}}\delta_{ll^{\prime}}\\
 \sqrt{\frac{k_1}{k_2}}S^{Jl}_{12}\delta_{JJ^{\prime}}\delta_{MM^{\prime}}\delta_{ll^{\prime}}
 &
 U_{2;JMl;J^{\prime}M^{\prime}l^{\prime}}
 -S^{Jl}_{22}\delta_{JJ^{\prime}}\delta_{MM^{\prime}}\delta_{ll^{\prime}}
 \ea
 \right|
 =0\;.
 \label{luscher_formula1}
 \ee
 \ew

 \subsubsection{L\"uscher's formula with definite cubic symmetry}

 In the sector with definite cubic symmetry, the basis of the representations are
 $|\Gamma,\xi;J,l,n\rangle$, where $\Gamma$ labels the symmetry sector (the irreducible
 representation of the cubic group), $\xi$
 runs from $1$ to the number of dimension for the irreducible representation, and
 $n$ runs from $1$ to the multiplicity of the irreducible representation. Then
 it can be expressed by the linear combination of $|JMl\rangle$, where the
 matrix $\mathcal{M}$ is diagonal with respect to $\Gamma$ and $\xi$ by Schur's lemma.
 In the symmetry sector $\Gamma$, the general formula~(\ref{luscher_formula1})
 is reduced into:
 \be
 \left|
 \ba
 [c]{cc}
 U_1(\Gamma)-S^{Jl}_{11} & \sqrt{\frac{k_{2}}{k_{1}}}S^{Jl}_{21}\\
 \sqrt{\frac{k_{1}}{k_{2}}}S^{Jl}_{12} & U_2(\Gamma)-S^{Jl}_{22}%
 \ea
 \right|   =0\;,
 \ee
 where $U_i(\Gamma)$ represent a linear transformations of the space
 $\mathcal{H}_{\Lambda}(\Gamma)$ with $J\leq \Lambda$.
 In terms of matrix element, assuming that
 the irreps $\Gamma$ appears only once, we will label
 them as $U_i(\Gamma)_{J^{\prime}l^{\prime};Jl}$.
 To write out a more explicit formula,
 we should consider the definite cubic symmetries. For the
 case of half-integer total momentum $J$, we need to consider the double cover
 group of $O$ denoted by $O^{D}$, which contains $48$ elements and can be
 divided into $8$ conjugate classes: $A_{1}$, $A_{2}$, $E$, $T_{1}$, $T_{2}$,
 $G_{1}$, $G_{2}$, $H$. For instance, $\ $for $J=\frac{1}{2}$, $\frac{3}{2}$,
 $\frac{5}{2}$, $\frac{7}{2}$, $\Lambda=4$, the decomposition into irreducible
 representation are given by $\frac{1}{2}=G_{1}$, $\frac{3}{2}=H$,
 $\frac{5}{2}=H\bigoplus{G_{2}}$,
 $\frac{7}{2}=H\bigoplus{G_{1}}\bigoplus{G_{2}}$ respectively~\cite{Basak:2005ir}.

 Now we focus on the $G_{1}$ and $G_{2}$ sector. In $G_{1}$ sector, there is a
 mixing between $J=\frac{1}{2}$ ($l=0,1$) and $J=\frac{7}{2}(l=3,4)$. If we
 neglect this mixing, then there is only mixing within $J=1/2$
 between $l=0$ and $l=1$, i.e. between $s$ wave and $p$ wave. In this case,
 L\"uscher's formula takes the following form
 \bw
 \be
 \left|
 \ba
 {cccc}
 U_{1;\frac{1}{2},0;\frac{1}{2},0}
 -S^{\frac{1}{2}0}_{11}
 &   \sqrt{\frac{k_2}{k_1}}S^{\frac{1}{2}0}_{21}
 & U_{1;\frac{1}{2},0;\frac{1}{2},1} &0\\
 \sqrt{\frac{k_1}{k_2}}S^{\frac{1}{2}0}_{12}
 &
 U_{2;\frac{1}{2},0;\frac{1}{2},0}
 -S^{\frac{1}{2}0}_{22}
 &0 & U_{2;\frac{1}{2},0;\frac{1}{2},1}\\
 U_{1;\frac{1}{2},1;\frac{1}{2},0} &0
 &U_{1;\frac{1}{2},1;\frac{1}{2},1}
 -S^{\frac{1}{2}1}_{11}
 &   \sqrt{\frac{k_2}{k_1}}S^{\frac{1}{2}1}_{21}
 \\
 0 & U_{2;\frac{1}{2},1;\frac{1}{2},0}&
 \sqrt{\frac{k_1}{k_2}}S^{\frac{1}{2}1}_{12}
 &
 U_{2;\frac{1}{2},1;\frac{1}{2},1}
 -S^{\frac{1}{2}1}_{22}
 \ea
 \right|
 =0 \;.  \label{luscher_formula22}
 \ee
 \ew

 In $G_{2}$ sector,the situation is similar, and
 there exists a mixing between $d$-wave ($l=2$) and $f$-wave ($l=3$).
 L\"uscher's formula takes exactly the same form as Eq.~(\ref{luscher_formula22})
 except that all the labels of $l^{\prime},l=0,1$ are replaced by $l^{\prime},l=2,3$
 and $J=1/2$ by $J=5/2$.  \label{G2}

 \subsection{L\"{u}scher's formula in moving frames}
 \label{subsec:MF}

 In this subsection, we extend two-channel L\"{u}scher's formula
 that has been obtained in the previous subsection for meson-baryon
 scattering to moving frames (MF).
 This is necessary in some lattice applications since it provides more
 low-momentum modes for a given lattice. Although we will only focus
 on the case of meson-baryon scattering, similar steps can be
 followed in the case of hadron scattering with arbitrary spin.
 We will follow the notations in Ref.~\cite{Rummukainen:1995vs} below.

 We denote the four momenta of the two particles
 in the lab frame, which is the frame in which periodic boundary
 conditions are applied, by
 \be
 k=(E_1,\bk)\;,
 \;\;
 P-k=(E_2,\bP-\bk)\;,
 \ee
 with $E_1=\sqrt{\bk^2+m^{2}_1}$ and $E_2=\sqrt{(\bP-\bk)^2+m^{2}_2}$
 being the energies of the two particles in the lab frame and
 $m_{1}$ and $m_{2}$ being the mass value of the
 baryon and meson respectively.
 The total three momentum $\mathbf{P}\neq 0$ of the two-particle system  is
 quantized by the condition $\mathbf{P}=(2\pi/L)\mathbf{d}$
 with $\mathbf{d}\in \mathbbm{Z}^{3}$.
 The COM frame is then moving relative to the lab frame with a velocity
 \be
 \bv=\bP/(E_1+E_2)\;.
 \ee
 In the COM frame, the momenta of the two particles
 will be denoted by $\bk^{\ast}$ and $(-\bk^{\ast})$, respectively.
 $\bk^{\ast}$ is related to $\bk$  by conventional Lorentz boost:
 \be
 \bk^{\ast\parallel}=\gamma(\bk^\parallel-\bv E_1)\;,
 \bk^{\ast\perp}=\bk^\perp\;,
 \ee
 where the symbol $\perp$ and $\parallel$ designates
 the components of the corresponding vector perpendicular
 and parallel to $\bv$, respectively.
 For simplicity, the above relation is also denoted
 by the shorthand notation: $\bk^{\ast}=\vec{\gamma}\bk$.
 A similar transformation relation holds for the other particle.

 Let $\phi(\mathbf{r})$ represent the wave
 function of the system in lab frame where $\mathbf{r}$
 is the relative coordinate between the two particles.
 Next, we enclose the system in a cubic box with finite size $L>2R$ and apply
 periodic boundary conditions to $\phi(\mathbf{r})$.
 On the other hand, this wave function can be related to the COM wave function
 $\phi^{CM}(\mathbf{r})$ by a Lorentz transform. Periodic boundary conditions
 in $\phi(\bfr)$ then implies that $\phi^{CM}(\mathbf{r})$ fulfills the
 so-called $d$-periodic boundary condition~\cite{Rummukainen:1995vs,ZiwenFu2012,Davoudi:2011md}:
 \be
 \phi^{CM}(\mathbf{r})=e^{i\pi\alpha\mathbf{d}\cdot{\mathbf{n}}}
 \phi^{CM}(\mathbf{r}+\vec{\gamma}\mathbf{n}L)
 \ee
 with $\mathbf{n}\in{\mathbbm{Z}^{3}}$,
 $\alpha=1+(m_1^{2}-m_2^{2})/E^{\ast2}$,
 $E^{\ast}=\sqrt{m_1^{2}+\bk^{\ast2}}+\sqrt{m_2^{2}+\bk^{\ast2}}$ is the
 total energy in the center of mass
 frame and $\mathbf{d}=(L/2\pi)\mathbf{P}$.

 In two-channel scattering, the COM wave function of the system
 can be written as:
 \be
 \phi^{CM}(\mathbf{r})=\left(
 \ba
 [c]{c}%
 \phi^{CM}_{1}(\mathbf{r})\\
 \phi^{CM}_{2}(\mathbf{r})
 \ea
 \right)\;,
 \ee
 where the form of $\phi^{CM}_{i}(\mathbf{r})$ is the same as
 $\psi_{i}(\mathbf{r})$ given in Eq.~(\ref{eq:eins})
 in subsection~\ref{subsec:COM}.
 In the outer region, $\phi^{CM}_{i}(\mathbf{r})$ can be also expanded
 in terms of modified Green's function
 $G_{i;JMl}^{\mathbf{d}}(\mathbf{r};k_{i}^{\ast2})$,
 \bw
 \be
 \phi^{CM}(\mathbf{r})=\left(
 \ba
 [c]{c}%
 \overset{\infty}{\underset{l=0}{\sum}}\overset{l+\frac{1}{2}}
 {\underset{J=l-\frac{1}{2}}{\sum}}\overset{J}{\underset{M=-J}{\sum}}
 \nu_{1;JMl}G_{1;JMl}^{\mathbf{d}}(\mathbf{r};k_{1}^{\ast2})\\
 \overset{\infty}{\underset{l=0}{\sum}}\overset{l+\frac{1}{2}}
 {\underset{J=l-\frac{1}{2}}{\sum}}\overset{J}{\underset{M=-J}{\sum}}
 \nu_{2;JMl}G_{2;JMl}^{\mathbf{d}}(\mathbf{r};k_{2}^{\ast2})
 \ea
 \right)\;.
 \ee
 \ew
 Just like in COM frame, the corresponding Green's functions
 $G^{\mathbf{d}}_{i;JMl}(\textbf{r},k_i^{\ast2})$ are given by
 an analogous expansion,
 \bw
 \be
 \label{eq:Md}
 G^{\mathbf{d}}_{i;JMl}(\textbf{r},k_i^{\ast2})=
 \frac{(-1)^lk^{{\ast}(l+1)}_i}{4\pi}\left[Y_{JM}^{l\frac{1}{2}}(\mathbf{{\hat{r}}})n_l(k_i^{\ast}r)
 +\sum_{l^{\prime}=0}^\infty\sum_{J^{\prime}=l^{\prime}-\frac{1}{2}}^{l^{\prime}+\frac{1}{2}}\sum_{M^{\prime}=-J^{\prime}}^{J^{\prime}}
  \mathcal{M}^{\mathbf{d}}_{i;JMl;J^{\prime}M^{\prime}l^{\prime}}(k_i^{\ast2})
Y_{J^{\prime}M^{\prime}}^{l^{\prime}\frac{1}{2}}(\mathbf{{\hat{r}}})j_{l^{\prime}}(k_i^{\ast}r)
\right]\;.
 \ee
 \ew
 where the explicit expression for
 $\mathcal{M}_{i;JMl;J^{\prime}M^{\prime}l^{\prime}}^{\mathbf{d}}$ can be found
 in appendix A, e.g. Eq.~(\ref{eq:Md1}).
 We also define a unitary matrix as
 \be
 \label{eq:U_QM}
 U_{i;JMl;J^{\prime}M^{\prime}l^{\prime}}^{\mathbf{d}}%
 =\left(\frac{\mathcal{M}_i^{\mathbf{d}}+i}{\mathcal{M}_i^{\mathbf{d}}-i}\right)_{JMl;J^{\prime}M^{\prime}l^{\prime}}
 \;.
 \ee
 Then, following similar steps as in previous subsection, we can also obtain L\"{u}scher's
 formula in MF which takes exactly the same form as
 Eq.~(\ref{luscher_formula1}) except that all the matrix elements of
 $U_i$ are replaced by those of $U^{\mathbf{d}}_i$.

 In moving frames, to describe the scattering phase in definite symmetry sector, one should
 consider the cubic lattice space group $\mathcal{G}$ which is the semi-direct product of
 lattice translation group $\mathcal{T}$ and the double-covered cubic
 group $O^{D}$. The representation of $\mathcal{G}$ can be
 characterized by two indices: total three-momenta $\bP$ and
 a representation $\Gamma$ of the little group corresponding to momentum $\bP$.
 As examples, we will only discuss moving frames with total momentum
 $\bP=(\frac{2\pi}{L})\mathbf{e}_{3}$ (MF1) and
 with $\bP=(\frac{2\pi}{L})(\mathbf{e}_{1}+\mathbf{e}_{2})$
 (MF2)~\cite{XuFeng:2011} in the following.
 Together with MF1 and MF2, another moving frame with $\bP=(2\pi/L)(\mathbf{e}_{1}+\mathbf{e}_{2}+\mathbf{e}_{3})$ (MF3) has been
 discussed in Ref.~\cite{Gockeler:2012yj} for the case of single-channel (elastic) scattering.
 Strategies for the construction of lattice operators are also analyzed.
 In principle, these results  could be generalized to the case of
 multi-channels as well.

 In the case of MF1, the little group is $C_{8v}$, which has 7 conjugate classes:
 $A_{1}$, $A_{2}$, $B_{1}$, $B_{2}$, $E_{1}$, $E_{2}$, $E_{3}$. In $E_{1}$
 section, there is mixing between $J=\frac{1}{2}$ and $J=\frac{7}{2}$~\cite{Moore:2005dw}.
 If we neglect this mixing and assume an angular momentum cutoff of $\Lambda=4$,
 L\"{u}scher's formula becomes
 \bw
 \beq
 \left|
 \ba{cccc}
 U^{\mathbf{d}}_{1;\frac{1}{2},0;\frac{1}{2},0}
 -S^{\frac{1}{2}0}_{11}
 &   \sqrt{\frac{k_2^{\ast}}{k_1^{\ast}}}S^{\frac{1}{2}0}_{21}
 & U^{\mathbf{d}}_{1;\frac{1}{2},0;\frac{1}{2},1} &0\\
 \sqrt{\frac{k_1^{\ast}}{k_2^{\ast}}}S^{\frac{1}{2}0}_{12}
 &
 U^{\mathbf{d}}_{2;\frac{1}{2},0;\frac{1}{2},0}
 -S^{\frac{1}{2}0}_{22}
 &0 & U^{\mathbf{d}}_{2;\frac{1}{2},0;\frac{1}{2},1}\\
 U^{\mathbf{d}}_{1;\frac{1}{2},1;\frac{1}{2},0} &0
 &U^{\mathbf{d}}_{1;\frac{1}{2},1;\frac{1}{2},1}
 -S^{\frac{1}{2}1}_{11}
 &   \sqrt{\frac{k_2^{\ast}}{k_1^{\ast}}}S^{\frac{1}{2}1}_{21}
 \\
 0 & U^{\mathbf{d}}_{2;\frac{1}{2},1;\frac{1}{2},0}&
 \sqrt{\frac{k_1^{\ast}}{k_2^{\ast}}}S^{\frac{1}{2}1}_{12}
 &
 U^{\mathbf{d}}_{2;\frac{1}{2},1;\frac{1}{2},1}
 -S^{\frac{1}{2}1}_{22}
 \ea
 \right|
 &=&0 \;. \label{luscher_formula18}
 \eeq
 \ew

 In the case of MF2, the little group is $C_{4v}$, which has 5
 conjugate classes: $A_{1}$, $A_{2}$, $B_{1}$, $B_{2}$, $E$. When
 $\Lambda=4$, in $E$ section, there is  mixing among
 $J=\frac{1}{2},J=\frac{3}{2},J=\frac{5}{2}$ and
 $J=\frac{7}{2}$~\cite{Moore:2005dw}. The formula becomes quite
 complicated. However, if we only consider the case of
 $J=\frac{1}{2}$, the formula has the same form as Eq.~(\ref{luscher_formula18}).

 \section{Generalized L\"{u}scher's formula in quantum field theory}
 \label{sec:QFT}

 In this section, we will describe the generalized L\"uscher's
 formulae for meson-baryon scattering in
 quantum field theory. We will follow the strategy outlined in
 Refs.~\cite{Hansen:2012tf,Kim:2005gf}, see also Ref.~\cite{Christ:2005gi,Briceno:2012yi}.
 In what follows, we will first perform the discussion in a single channel case.
 It is then generalized to two-channel case in a straightforward manner. Then we
 compare what we obtained in quantum field theory with the results
 obtained in non-relativistic quantum mechanics in the previous section
 and show that they are equivalent, apart from possible corrections that
 are exponentially suppressed in the large volume limit.

 The discussion using relativistic quantum field theory has an
 advantage, namely the results are easily transformed to any frame, the COM frame or
 the lab frame (moving frame), both of which have been discussed in the previous section.
 In the single channel scenario, we denote the masses of the meson and the baryon as $m_1$ and
 $m_2$, respectively. The total four momentum of the two-particle system
 is denoted as $P=(E,\bP)$ and the quantities
 in the COM frame will be denoted by adding a * to the corresponding quantity.
 Thus, for example, the COM frame four momentum is denoted as $P^{\ast}=(E^{\ast}, \bzero)$.
 The individual three-momentum of the two particles will be denoted by $\bq^{\ast}$
 and $-\bq^{\ast}$, the magnitude of which ($q^{\ast2}=\bq^{{\ast}2}$) being:
 \be
 \label{eq:q_star_single}
 4q^{\ast2}=E^{\ast2}-2(m_{1}^{2}+m_{2}^{2})
 +\frac{(m_{1}^{2}-m_{2}^{2})^{2}}{E^{\ast2}}
 \;.
 \ee

\subsection{Single channel case}

 We start by deriving an expression from quantum field theory
 in the single channel case.
 We first use the method that have
 been studied in \cite{Hansen:2012tf,Kim:2005gf} to obtain the
 quantization condition, based on the periodic boundary conditions hence the total momentum
 being $\mathbf{P}=(2\pi/L){\mathbf{n}}$ with $(\mathbf{n}\in{\mathbbm{Z}^{3}})$.

 The basic idea in Ref.~\cite{Hansen:2012tf,Kim:2005gf} is the following:
 two-particle spectrum of the system in a finite box
 can be determined from the poles of an appropriate correlation function
 in the energy plane. Thus, one defines
 \be
 \label{eq:CP_definition}
 C(P)=\int_{L;x}e^{i(-\mathbf{P}\cdot \mathbf{x}+Ex^{0})}\langle0|\sigma(x)\sigma^\dagger(0)|0\rangle\;,
 \ee
 where $P=(E,\mathbf{P})$ is the total
 four-momentum of the two-particle system.
 The generic interpolating operator $\sigma(x)$ is chosen to have an overlap with the
 two-particle states that we are interested in (in our case, a meson and a baryon) and
 $\int_{L;x}=\int_{L}d^{4}x$ stands for the space-time integration over the finite volume.
 Two-particle spectrum are exactly those poles in the $E$ plane of $C(P)$.

 \begin{figure}[h]
 \centering \includegraphics[width=0.6\textwidth]{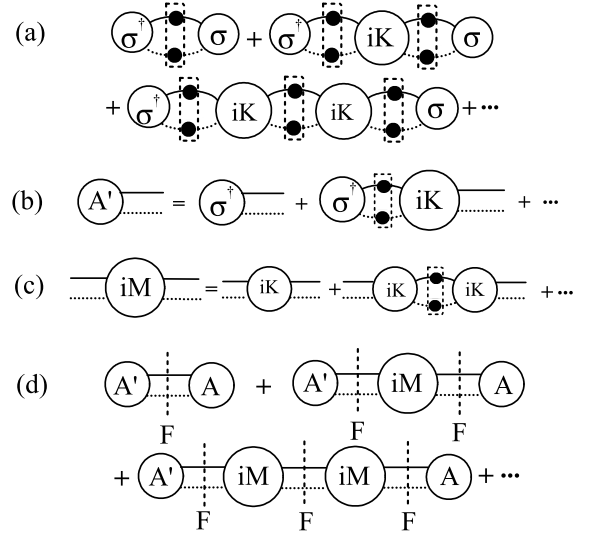}
 \caption{Horizontal dashed lines represent the meson propagators while
 the solid lines represent the baryon propagators.
 The corresponding full propagators are represented by
 dashed/solid lines with a black dot.
  The dashed rectangles indicate finite volume momentum
 sum/integrals.
 Each dashed vertical line indicates an insertion of
 the factor $F$ entering Eq.~(\ref{finite_corralator_1})
 which carries the volume dependence of interest.
 (a) The diagrams that build up the correlation function $C(P)$ defined
 in Eq.~(\ref{eq:CP_definition});
 (b) and (c) Diagrams that build up matrix element $A$
 and the scattering amplitude $iM$, respectively;
 (d) The resulting series for the subtracted
 correlator defined by Eq.~(\ref{eq:subtracted_C}).
 \label{fig:fig1}}%
 \end{figure}
 The correlation function $C(P)$ may be built from a series of
 contributions illustrated by diagrams in FIG.~\ref{fig:fig1}(a).
 In this figure, a solid line with a black dot stands for a full
 propagator of the baryon while a dashed line with
 a dot stands for that of the meson.
 A circle with a symbol $iK$ inside represents the Bethe-Salpeter kernel
 which consists of all amputated two-particle irreducible diagrams.
 A circle with a symbol $\sigma$ or $\sigma^\dagger$ denotes
 the interpolating operators in Eq.~(\ref{eq:CP_definition}).
 Using $iK$ we may write the $C(P)$  in the following form:
 \beq
 \label{eq:sigma_BS}
 & &C(P)=\int_{L;q}\sigma_{q}\left[Z_{1}\Delta_{1}Z_{2}\Delta_{2}\right]_{q}
 \sigma_{q}^\dagger\nonumber\\
 &+&
 \int_{L;q,q^{\prime}}\sigma_{q}\left[Z_{1}\Delta_{1}Z_{2}\Delta_{2}\right]_{q}
 iK_{q,q^{\prime}}\left[Z_{1}\Delta_{1}Z_{2}\Delta_{2}\right]_{q^{\prime}}\sigma_{q^{\prime}}^{\dagger}+\cdots\nonumber\\
 \eeq
 where we have adopted the shorthand notation for the two-particle propagators
 \be
 \left[Z_{1}\Delta_{1}Z_{2}\Delta_{2}\right]_{q}
 =[Z_{1}(q)\Delta_{1}(q)]\cdot[Z_{2}(q)\Delta_{2}(P-q)]
 \;.
 \ee
 Denoting the interpolating field for the baryon and the meson by
 $\phi_{1}(x)$ and $\phi_2(x)$, respectively,
 the full propagators appearing in the above equations read
 \beq
 Z_{i}(q)\Delta_{i}(q)&=&\int{d^{4}x}e^{iq\cdot x}\langle
 \phi_{i}(x)\phi_{i}(0)\rangle\;,\;
 \\
 \Delta_{1}(q)&=&\frac{i(q^{\mu}\gamma_{\mu}+m_{1})}
 {q^{2}-m_{1}^{2}+i\varepsilon}
 \;,\\
 \Delta_{2}(q) &=&\frac{i}{q^{2}-m_{2}^{2}+i\varepsilon}\;.
 \eeq
 The factors $Z_{1}(q)$ and $Z_{2}(q)$ are the corresponding
 dressing functions for the baryon and the meson, respectively.
 In Eq.~(\ref{eq:sigma_BS}), two-particle intermediate states
 are summed/integrated in a manner that is appropriate for the finite volume,
 namely,
 \be
 \int_{L,q}=\frac{1}{L^{3}}\sum_{\mathbf{q}}\int \frac{dq^0}{2\pi}
 \;.
 \ee
 The kernel $iK$ is related to the Bethe-Salpeter kernel $BS$ as
 \be
 iK_{q,q^{\prime}}= iBS(q,P-q,-q^{\prime},-P+q^{\prime})\;,
 \ee
 with $BS$ the sum of all amputated two-particle scattering diagrams which are
 two-particle-irreducible. Finally the factors $\sigma_q$ and
 $\sigma^\dagger_q$ denote the coupling of the interpolating operator
 to the two-particle states.

 Normally the interpolating operator
 is the product of two interpolating operators for the two hadrons
 being considered, e.g. $\sigma(x)=\phi_1(x)\phi_2(x)$.
 In order to have an overlap with the desired states,
 interpolating operators are usually designed to carry definite $J^{PC}$
 quantum numbers. Therefore, without loss of generality, we assume
 that $\sigma(x)$ carries definite parity.
 In our case, the operator $\sigma(x)$ must carry
 Dirac indices as $\phi_1(x)$ and since $\gamma_0$ is the Dirac matrix
 responsible for parity transformation for Dirac spinors, we assume that
 $\sigma(x)$ commute with $\gamma_0$. In other words, in what
 follows, $\gamma_0$ may be considered as a $c$-number rather
 than a matrix in Dirac space.

 The kernel $iK$ and the propagator dressing functions
 $Z_{1,2}(q)$ have only exponentially suppressed dependence on
 the box size $L$. Such dependence is always assumed to be small
 and negligible in the large volume limit.
 The dominant power-law volume dependence enters
 through the discrete momentum sums in the two-particle loops in
 Eq.~(\ref{eq:sigma_BS}), the details of these calculations are outlined in
 appendix~\ref{appendix:loops}, see Eq.~(\ref{loop_2}).
 Basically, each summation/integration can be split into two parts:
 \be
 I=I_\infty+I_{FV}\;,
 \ee
 where $I_\infty$ designates the infinite volume result of
 the loop integral and $I_{FV}$ contains the finite volume corrections.
 For each of the loop summations in FIG.~\ref{fig:fig1}(a), if we
 take $I_{\infty}$ in each loop, we then recover the infinite volume correlator
 $C^{\infty}(P)$. As shown in appendix~\ref{appendix:loops}, $C^{\infty}(P)$ does not
 contain the two-particle poles that we are looking for.
 The part of interest is the finite volume correction:
 \be
 \label{eq:subtracted_C}
 C^{FV}(P)=C(P)-C^{\infty}(P)\;,
 \ee
 where $C^{FV}(P)$ and $C^{\infty}(P)$ are the finite and infinite volume correlation functions,
 respectively. Diagrammatically, $C^{FV}(P)$ is obtained by
 keeping at least one insertion of $I_{FV}$ in the two-particle loops
 in Eq.~(\ref{eq:sigma_BS}).
 Let $F$ be the kinematic factor associated with the factor of $I_{FV}$.
 These contributions are shown in Fig.~\ref{fig:fig1}(d),
 leading to the following general result:
 \begin{align}
 C^{FV}  &  =-AFA^{\prime}+AF(iM)FA^{\prime}+\ldots\nonumber\\
 &  =-AF\frac{1}{1+iMF}A^{\prime} \;.
  \label{finite_corralator_1}
 \end{align}
 The finite volume correction $C^{FV}$ contains the two-particle poles we are looking for.

 The factor $F$ is discussed in some detail in appendix~\ref{appendix:loops},
 see Eq.~(\ref{field_F3}). It arises from the integration/summation of
 the intermediate states represented by the loop diagram, see
 Eq.~(\ref{loop_sum_1}). Due to the fermion propagator, the factor
 $F$ is in principle a matrix in Dirac space, as shown in
 Eq.~(\ref{eq:def_of_CD}). However, it only contains
 the matrix $\gamma_0$, which is the matrix responsible for the transformation of a
 Dirac spinor under parity. Since in practical
 applications we normally choose the interpolating operators $\sigma(x)$ to
 carry definite parity quantum number, this means that $\gamma_0$
 can be replaced by its eigenvalues: $\pm1$. In other words,
 we can simply treat the function $C(\bk)$ defined in Eq.~(\ref{eq:def_of_CD})
 as $c$-numbers. Thus, in the COM frame, we simply define:
 \be
 F=C(\bq^{\ast})\tilde{F}\;,
 M=C(\bq^{\ast})^{-1}\tilde{M}\;.
 \ee
 with this, Eq.~(\ref{finite_corralator_1}) becomes
  \begin{align}
 C^{FV}  &  =-AFA^{\prime}+AF(iM)FA^{\prime}+\ldots\nonumber\\
 &  =-AC(\mathbf{q^{\ast}})\tilde{F}A^{\prime}+AC(\mathbf{q^{\ast}})\tilde{F}(i\tilde{M})\tilde{F}A^{\prime}+\ldots\nonumber\\
 &  =-AC(\mathbf{q^{\ast}})\tilde{F}\frac{1}{1+i\tilde{M}\tilde{F}}A^{\prime} \;.
 \label{finite_corralator_2}
 \end{align}

 The factors $A$ and $A^{\prime}$ appearing in Eq.~(\ref{finite_corralator_1})
 may be expressed by appropriate
 matrix elements of the interpolating operator, as shown in Fig.~\ref{fig:fig1}(b).
 In COM frame the amplitudes $A$ and $A^{\prime}$ now read,
 \be
 \label{eq:A_s}
 \left\{ \begin{aligned}
 A(\mathbf{\hat{k}}^{\ast})&=\langle0|\sigma(0)
 |\mathbf{k}^{\ast},-\mathbf{k}^{\ast};\mbox{\rm in}\rangle_{|\mathbf{k}^{\ast}|=q^{\ast}}
 \\
 A^{\prime}(\mathbf{\hat{k}}^{\ast})&=\langle \mathbf{k}^{\ast},-\mathbf{k}^{\ast};\mbox{\rm out}|\sigma^\dagger(0)|0\rangle_{|\mathbf{k}^{\ast}|=q^{\ast}}\end{aligned} \right.
  \;.
 \ee
 Both of these two amplitudes can be viewed as two-component spinor
 in spin space.
 It is more useful to express the abstract formula~(\ref{finite_corralator_2})
 in angular momentum basis. For this purpose,
 we expand the matrix elements in terms of spin spherical harmonics
 defined in Eq.~(\ref{eq:spin_spherical_harmonics_def}):
 \be
 \label{eq:A_s_expand}
 \left\{ \begin{aligned}
 A({\mathbf{\hat{k}}^{\ast}})&=\sqrt{4\pi}A_{JMl}Y_{JM}^{l\frac{1}{2}\dagger}(\mathbf{\hat{k}}^{\ast})
 \\
 A^{\prime}(\mathbf{\hat{k}^{\ast}})&=\sqrt{4\pi}A^{\prime}_{JMl}Y_{JM}^{l\frac{1}{2}}(\mathbf{\hat{k}}^{\ast})
 \end{aligned} \right.\;,
 \ee
 where summation over repeated indices (i.e. $J$, $M$ and $l$) are understood.

 Similarly, working in states with definite parity,
 $\tilde{F}$ and $\tilde{M}$ are viewed as $2\times 2$ matrices in spin
 space, which can be expanded as well,
 \bw
 \beq
 \tilde{F}(\mathbf{\hat{k}}^{\ast},\mathbf{\hat{k}^{\ast \prime}})&=&\frac{-1}{4\pi}\tilde{F}_{JMl;J^{\prime}M^{\prime}l^{\prime}}
 Y_{JM}^{l\frac{1}{2}}
 (\mathbf{\hat{k}^{ \ast}})
 Y_{J^{\prime}M^{\prime}}^{l^{\prime}\frac{1}{2}{\dagger}}(\mathbf{\hat{k}^{\ast \prime}})\;,
 \eeq
 \beq
 \tilde{M}(\mathbf{\hat{k}}^{\ast},\mathbf{\hat{k}^{\ast \prime}})&=&{4\pi}\tilde{M}_{JMl;J^{\prime}M^{\prime}l^{\prime}}
 Y_{JM}^{l\frac{1}{2}}
 (\mathbf{\hat{k}^{\ast}})Y_{J^{\prime}M^{\prime}}^{l^{\prime}\frac{1}{2}{\dagger}}(\mathbf{\hat{k}^{\ast \prime}})\;.
 \eeq
 \ew
 Here the notation $Y_{JM}^{l\frac{1}{2}}
 (\mathbf{\hat{k}^{\ast}})Y_{J^{\prime}M^{\prime}}^{l^{\prime}\frac{1}{2}{\dagger}}(\mathbf{\hat{k}^{\ast \prime}})$
 stands for the direct product of two spin spherical harmonics.
 Then the abstract formula Eq.(\ref{finite_corralator_2}) remains valid except that
 all implicit indices are in angular momentum space.
 For example:
 \bw
 \beq
 (AC(\mathbf{q^{\ast}})\tilde{F}\tilde{M}\tilde{F}A^{\prime})_{J_1M_1l_1;J_4M_4l_4} &=&A_{J_{1}M_{1}l_{1}}C(\mathbf{q^{\ast}})\tilde{F}_{J_{1}M_{1}l_{1};J_{2}M_{2}l_{2}}
 \tilde{M}_{J_{2}M_{2}l_{2};J_{3}M_{3}l_{3}}\tilde{F}_{J_{3}M_{3}l_{3};J_{4}M_{4}l_{4}}A^{\prime}_{J_{4}M_{4}l_{4}}\;,
 \eeq
 \ew
 where the matrix $\tilde{F}_{JMl;J^{\prime}M^{\prime}l^{\prime}}$ is given by:
 \begin{align}
 \tilde{F}_{JMl;J^{\prime}M^{\prime}l^{\prime}}  &  =\frac{q^{\ast}}
 {8\pi{E^{\ast}}}(\delta_{JJ^{\prime}}\delta_{MM^{\prime}}\delta_{ll^{\prime}}
 +iF_{JMl;J^{\prime}M^{\prime}l^{\prime}}^{FV})
 \label{F_definition}
 \end{align}
 with $\tilde{F}_{JMl;J^{\prime}M^{\prime}l^{\prime}}^{FV}$  given
 in Eq.~(\ref{field_F1}) in the appendix.

 The matrix $\tilde{M}$ in Eq.~(\ref{finite_corralator_2}) is the
 scattering amplitude illustrated in FIG.~\ref{fig:fig1}(b) and FIG.~\ref{fig:fig1}(c)
 which can be related to the non-relativistic quantum mechanical
 scattering matrix $M^{(NR)}$.
 Following the discussion as in Ref.~\cite{luscher91:finiteb},
 the relation is found to be
 \beq
 \tilde{M}_{JMl;J^{\prime}M^{\prime}l^{\prime}}&=&
 8\pi{E^{\ast}}M^{(NR)}_{JMl;J^{\prime}M^{\prime}l^{\prime}}
 \;.
 \label{field_scattering_amplitude}
 \eeq
 where $M^{(NR)}_{JMl;J^{\prime}M^{\prime}l^{\prime}}$  given by
 Eq.~(\ref{define_M_nr}) in appendix~\ref{appendix:spin_scattering}.

 The quantization condition can now be obtained
 for $C^{FV}(P)$ which manifest itself as a series of two-particle poles.
 This means that the matrix between
 $A$ and $A^{\prime}$ must have divergent eigenvalues hence
 satisfy the following condition:
 \be
 \det(1+i\tilde{M}\tilde{F})=0 \;.
 \label{luscher_formula_3}
 \ee
 This is the so-called quantization condition for the
 two-particle poles in a finite box.
 In order to compare this with the conventional
 L\"{u}scher's formula obtained by non-relativistic approach, we substitute
 the definitions of $\tilde{M}_{JMl;J^{\prime}M^{\prime}l^{\prime}}$ and $\tilde{F}_{JMl;J^{\prime}M^{\prime}l^{\prime}}$ in
 Eq.~(\ref{field_scattering_amplitude}) and Eq.~(\ref{F_definition})
 into Eq.~(\ref{luscher_formula_3}). We then obtain the following result:
 \beq
 \det(\tan{\delta_{Jl}}F_{JMl;J^{\prime}M^{\prime}l^{\prime}}^{FV}
 -\delta_{JJ^{\prime}}\delta_{MM^{\prime}}\delta_{ll^{\prime}})&=&0 \;.\label{luscher_formula10}
 \eeq

 Using an equality which will be shown in appendix~\ref{appendix:loops},
 namely Eq.~(\ref{equality_1}), it is seen that $F$ is related to
 the matrix $\mathcal{M}^{\mathbf{d}}$ up to a possible phase.
 In the case of strong interaction, the phase factor does not enter,
 since the Hamiltonian of the system conserves parity and there can be no scattering
 connecting $l=J+\frac{1}{2}$ with $l=J-\frac{1}{2}$.
 So the scattering matrix $S^{Jl}=\exp(2i\delta_{Jl})$ must be a diagonal matrix.
 It is then verified that Eq.~(\ref{luscher_formula10}) reduces to
 the conventional single-channel L\"uscher's formula in a moving frame
 discussed in subsection~\ref{subsec:MF}.

\subsection{Two-channel case}

 In this subsection, we generalize the results of a single channel
 case to the two-channel case. The formalism is the same as in the single channel case
 except that we need two interpolating operators $\sigma_j(x)$ with  $j=1,2$ denoting two
 different channels. In this case, as shown in Fig.~\ref{fig:fig1},
 the two-point function has the following form,
 \bw
 \beq
 C(P)=\int_{L;q}\sigma_{j;q}[Z_{1}\Delta_{1}
 Z_{2}\Delta_{2}]_{jg;q}\sigma_{g;q}^\dagger
 +\int_{L;q,q'}\sigma_{j;q}[Z_{1}\Delta_{1}Z_{2}\Delta_{2}]_{jg;q}
 iK_{gh;q,q'}
 [Z_{1}\Delta_{1}Z_{2}\Delta_{2}]_{hr;q'}\sigma_{r;q^{\prime}}^\dagger+\cdot\cdot\cdot
 \label{eq}
 \eeq
 \ew
 Here indices $g$, $h$, and $r$ also refer to the channel
 and take the value $1$ or $2$. In Eq.(\ref{eq}), we have also utilized
 the following shorthand notations:
 \be
 \left[Z_{1}\Delta_{1}Z_{2}\Delta_{2}\right]_{jg;q}
 =\delta_{jg}[Z_{j;1}\Delta_{j;1}(q)][Z_{g;2}\Delta_{g;2}(P-q)]
 \;,
 \ee
 with the definitions
 \be
 Z_{j;I}\Delta_{j;I}(q)=\int{d^{4}x}e^{iq\cdot x}\langle
 \phi_{j;I}(x)\phi_{j;I}(0)\rangle \;,
 \ee
 \be
 \Delta_{j;1}(q)=\frac{i(q^{\mu}\gamma_{\mu}+m_{j;1})}
 {q^{2}-m_{j;1}^{2}+i\varepsilon}\;,
 \ee
 \be
 \Delta_{j;2}(q)=\frac{i}{q^{2}-m_{j;2}^{2}+i\varepsilon}\;.
 \ee
 So we have two indices for the propagators and their dressing
 functions: $\Delta_{j;I}$ (and the corresponding mass values $m_{j;I}$ which enters
 the propagator) and $Z_{j;I}$.
 The first index $j=1,2$ designates two different scattering channels.
 The second index $I=1,2$
 now denotes particle types: $I=1$ for the baryon and $I=2$ for the meson.

 The kernel $K_{gh}$ is again related to the Bethe-Salpeter kernel
 $BS_{gh}$ which now becomes a matrix in channel space.
 The matrix elements of interpolating fields $\sigma_j$
 become vectors in channel space.
 We have also assumed that $\sigma_j$ carry definite parity quantum numbers.
 So similar to Eq.~(\ref{eq:A_s}), we have,
 \be
 \left\{ \begin{aligned}
  A_{j}(\mathbf{\hat{k}^{\ast}})&=\langle0|\sigma_j(0)|\mathbf{k}^{\ast},-\mathbf
 {k}^{\ast};j;\mbox{\rm in}\rangle_{|\mathbf{k}^{\ast}|=q_{j}^{\ast}}\\
 A_{j}^{\prime}(\mathbf{\hat{k}^{\ast}})&=\langle
 \mathbf{k}^{\ast},-\mathbf{k}^{\ast};j;\mbox{\rm out}|\sigma^\dagger_j(0)|0
 \rangle_{|\mathbf{k}^{\ast}|=q_{j}^{\ast}}
 \end{aligned} \right.\;.
 \ee
 In angular momentum basis, expansion~(\ref{eq:A_s_expand}) becomes
 \be
 \left\{ \begin{aligned}
  A_{j}(\mathbf{\hat{k}^{\ast}})&=\sqrt{4\pi}A_{j;JMl}Y_{JM}^{l\frac{1}{2}{\dagger}}
 (\mathbf{\hat{k}^{\ast}})\\
 A_{j}^{\prime}(\mathbf{\hat{k}^{\ast}})&=\sqrt{4\pi}A^{\prime}_{j;JMl}Y_{JM}^{l\frac{1}{2}}(\mathbf{\hat{k}^{\ast}})
 \end{aligned} \right.\;.
 \ee
 The factors $\tilde{F}$ and $\tilde{M}$ entering the
 quantization condition~(\ref{luscher_formula_3})
 in the previous subsection have both become $2\times 2$
 matrices in channel space. We may expand them
 in terms of spin spherical harmonics,
 with the coefficients $\tilde{F}$ being diagonal in channel space:
 \beq
 \tilde{F}_{ij;JMl;J^{\prime}M^{\prime}l^{\prime}}&=&\delta_{ij}\tilde{F}_{i;JMl;J^{\prime}M^{\prime}l^{\prime}}\;.
 \label{field_scattering_F}
 \eeq
 The diagonal element is given by
 \be
 \label{eq:F_FV}
 \tilde{F}_{i;JMl;J^{\prime}M^{\prime}l^{\prime}}=\frac{q^{\ast}}
 {8\pi{E^{\ast}}}(\delta_{JJ^{\prime}}\delta_{MM^{\prime}}\delta_{ll^{\prime}}
 +iF_{i;JMl;J^{\prime}M^{\prime}l^{\prime}}^{(FV)})\;.
 \ee
 The scattering matrix $M$, however, is not diagonal in channel space.
 It is still related to the
 non-relativistic scattering amplitude via,
 \beq
 \tilde{M}_{ij;JMl;J^{\prime}M^{\prime}l^{\prime}}&=&
 8\pi{E^{\ast}}M^{(NR)}_{ij;JMl;J^{\prime}M^{\prime}l^{\prime}}\;,
 \label{field_scattering_amplitude_1}
 \eeq
 where  $M^{(NR)}_{ij;JMl;J^{\prime}M^{\prime}l^{\prime}}$ is defined in
 Eq.~(\ref{define_multi_M_nr}).
 According to FIG.~\ref{fig:fig1}1 (d),
 Eq.~(\ref{finite_corralator_2}) still holds
 except that the quantities involved have all become matrices or vectors in
 channel space.
 The poles in $C^{FV}$ still yield the desired
 quantization condition~(\ref{luscher_formula_3}) with the
 understanding that both $\tilde{M}$ and $\tilde{F}$ have now become $2\times 2$
 matrices in channel space.

 We are now in a position to write out the quantization condition
 to a form that is comparable to the conventional multi-channel
 L\"uscher formula obtained in the previous section.
 For this purpose, we define a matrix $U^{(FV)}_i$ using
 the matrix $F^{(FV)}_i$ in Eq.~(\ref{eq:F_FV}),
 \be
 \label{eq:U_QFT}
 U_{i;JMl;J^{\prime}M^{\prime}l^{\prime}}^{(FV)}
 =\left(  \frac{F_i^{(FV)}+i}{F_i^{(FV)}-i}\right)_{JMl;J^{\prime}M^{\prime}l^{\prime}}\;.
 \ee
 Now consider an arbitrary  moving frame.
 As we will show in appendix~\ref{appendix:loops}, see
 Eq.~(\ref{equality_2}),
 the matrix element $F^{(FV)}_{i;JMl;J'M'l'}$
 is in fact related to the corresponding matrix element $\mathcal{M}^{\mathbf{d}}_{i;JMl;J^{\prime}M^{\prime}l^{\prime}}$
 that have appeared in our discussion in subsection~\ref{subsec:MF},
 c.f. Eq.~(\ref{eq:Md}):
 \be
 F^{(FV)}_{i;JMl;J^{\prime}M^{\prime}l^{\prime}}=i^{l-l^{\prime}}\mathcal{M}^{\mathbf{d}}_{i;JMl;J^{\prime}M^{\prime}l^{\prime}}\;.
 \ee
 This expression is valid up to terms that are exponentially suppressed in the
 large volume limit. Note that if the Hamiltonian conserves parity
 such that scattering only occurs for $l=l^{\prime}$, the two matrices
 become identical. Substituting the definitions of $\tilde{M}_{ij;JMl;J^{\prime}M^{\prime}l^{\prime}}(\ref{field_scattering_amplitude_1})$
 and  $\tilde{F}_{ij;JMl;J^{\prime}M^{\prime}l^{\prime}}(\ref{field_scattering_F})$
 to the quantization condition Eq.~(\ref{luscher_formula_3}),
 we arrive at the following result:
 \begin{widetext}
 \beq
 \left \vert
 \begin{array}
 [c]{cc}
 (F_{1;JMl;J^{\prime}M^{\prime}l^{\prime}}^{(FV)}+i)-S_{11}^{Jl}(F_{1;JMl;J^{\prime}M^{\prime}l^{\prime}}^{(FV)}-i) & \sqrt{\frac{q_{2}^{\ast}}{q_{1}^{\ast}}}
 S_{21}^{Jl}(F_{2;JMl;J^{\prime}M^{\prime}l^{\prime}}^{(FV)}-i)\\
 \sqrt{\frac{q_{1}^{\ast}}{q_{2}^{\ast}}}S_{12}^{Jl}(F_{1;JMl;J^{\prime}M^{\prime}l^{\prime}}^{(FV)}-i) & (F_{2;JMl;J^{\prime}M^{\prime}l^{\prime}}^{(FV)}+i)
 -S_{22}^{Jl}(F_{2;JMl;J^{\prime}M^{\prime}l^{\prime}}^{(FV)}-i)
 \end{array}
 \right \vert &=&0\;.\label{luscher_formula30}
 \eeq
 \end{widetext}
 Comparing (\ref{luscher_formula28}) with
 (\ref{luscher_formula30}), we find that
 the forms of these formulae are completely
 equivalent although we have used different methods to obtain them.
 This conclusion is valid up to terms which vanish exponentially
 with the box size.

 \section{Discussions and conclusions}
 \label{sec:conclude}

 In this paper, we have generalized L\"uscher's formula to the
 case of multi-channel two-particle
 (one spinless, one spin-$1/2$) scattering in a cubic box.
 The generalization was done using both non-relativistic quantum
 mechanics and quantum field theory. We verified that, up to terms that are
 exponentially suppressed in the large volume limit, both methods
 yield compatible results.

 Although we only consider the case of meson-baryon scattering
 in this paper, using similar techniques, it should not be too difficult to
 generalize the results to the case of scattering between hadrons with other
 spin configurations. In fact, using similar notations as in
 Refs.~\cite{Bernard:2008ax,Ishizuka:2009bx}, one should be able to obtain
 corresponding formulae suitable
 for nucleon-nucleon multi-channel scattering.
 Another interesting direction is the corresponding formulae in
 a box with different boundary conditions, which turns out to be useful
 for practical reasons.
 In particular, the formulae obtained in this paper can readily be generalized for
 anti-periodic or twisted boundary conditions.

 A unique feature of the multi-channel scattering L\"uscher's formula which differs from
 that in the single-channel case is that, the corresponding equation is not a one-to-one
 relation between the energy and the corresponding scattering parameters.
 Therefore, even if one can construct
 appropriate correlation functions to obtain the two-particle energy $E$ in lattice
 simulations, L\"uscher's formulae only sets up
 constraints among the energy $E$ and the $S$-matrix parameters $S^{Jl}_{ij}(E)$.
 Further physical inputs are needed to really pin down these
 scattering parameters in a multi-channel scenario.

 Finally, let us discuss some possible applications of L\"{u}scher's
 formula in multi-channel scattering. One typical example is the antikaon-nucleon scattering.
 The scattering amplitude of antikaon-nucleon is of fundamental importance in
 the study of $\Lambda$(1405) resonance
 which just exceeds the scattering threshold.
 There is a strong coupling between
 $\bar{K}$N and $\Sigma \pi$ channels when the energy exceeds $\bar{K}$N threshold. It
 becomes a problem for two-particle scattering,
 one with spin $\frac{1}{2}$ and one with spin $0$ in two channels.
 In Ref.~\cite{Lage:2009zz}, the authors used two channel
 Lippmann-Schwinger equation to study the problem.
 In Ref.~\cite{MartinezTorres:2012yi}, the same problem was
 addressed using unitarized chiral perturbation theory.
 In principle, this problem can also be studied using L\"uscher's formula
 in lattice QCD simulations, although that requires more data than
 we currently can acquire.
 Another example is the $DN$ and $\pi \Sigma_{c}$ coupled channels
 scattering studied in Ref.~\cite{Xie:2012pi}.

 To summarize, in this paper we generalize L\"{u}scher's formula to the case of particles
 with spin, beyond the inelastic threshold in COM frame and MF respectively.
 Using a quantum mechanical model and quantum field theory,
 a relation between the energy of the two-particle system and scattering matrix
 elements is found. It is verified that the two methods yields the same result
 if we neglect terms that are exponentially suppressed in the large volume limit.
 Although we focus on the case of scattering between two particles: one with spin
 $0$ and the other with spin $\frac{1}{2}$, we do not see any
 essential difficulties to generalize the situation to any spin
 with any number of channels. We hope that these relations will
 help us to study multi-channel hadron scattering between particles with spin
 using lattice QCD in the future.

\section*{Acknowledgements}

 This work is supported in part by the
 National Science Foundation of China (NSFC) under the project
 No. 10835002 and No.11021092.
 It is also supported in part by the DFG and the NSFC (No.11261130311) through funds
 provided to the Sino-Germen CRC 110 ``Symmetries and the Emergence
 of Structure in QCD''.

\appendix

\section{Single-channel spin-dependent scattering}
\label{appendix:spin_scattering}

 In this appendix, we first briefly review single
 channel scattering of a particle with spin in quantum mechanics
 in infinite volume. After that, single channel L\"uscher's formulae
 are collected which can be
 found in Refs.~\cite{Bernard:2008ax,Ishizuka:2009bx,Rummukainen:1995vs,ZiwenFu2012}.
 Similar formulae in the multi-channel case are listed afterwards at the end of
 this appendix.

 To avoid inessential complications, we shall only discuss the case of two stable
 particles with spin $0$ and spin $\frac{1}{2}$ in COM frame.
 In non-relativistic quantum mechanics, after factoring out the center of mass motions,
 the asymptotic form of the wave function is:
 \be
 \psi_s(\mathbf{r})=\chi^{\frac{1}{2}}_{s}e^{i\mathbf{k\cdot{r}}}
 +\sum_{s^{\prime}}\chi^{\frac{1}{2}}_{s^{\prime}}M^{(NR)}_{s^{\prime}s}(\mathbf{\hat{k}\cdot\hat{r}})\frac{e^{ikr}}{r}\;.
 \label{wavefunction10}
 \ee
 This form has the property that, in the remote past,
 it reduces to an incident plane wave with prescribed quantum numbers (linear momentum and
 spin).
 In Ref.~\cite{Newton},
 the scattering amplitude between particles with spin 0 and
 spin $\frac{1}{2}$ is given by,
 \bw
 \beq
 M^{(NR)}_{s^{\prime}s}(\mathbf{\hat{k}\cdot{\hat{r}}})
 &=&\frac{4\pi}{2ik}
 \sum_{l=0}^\infty\sum_{J=l-\frac{1}{2}}^{l+\frac{1}{2}}\sum_{M=-J}^{J}
 (S^{Jl}-1)\mathfrak{Y}_{JMls^{\prime}}(\mathbf{\hat{r}})\mathfrak{Y}_{JMls}^{\ast}(\mathbf{\hat{k}})\;,
 \label{scattering_amplitude_1}
 \eeq
 \ew
 where the auxiliary spin spherical harmonic
 function $\mathfrak{Y}_{JMls}(\mathbf{\hat{r}})$
 in spin-space is defined as
 \beq
 \mathfrak{Y}_{JMls}(\mathbf{\hat{r}})=i^{-l}\chi^{\frac{1}{2}\dagger}_{s}\cdot{Y^{l\frac{1}{2}}_{JM}(\mathbf{\hat{r}})}\;.
 \label{define_Y}
 \eeq
 The dot here indicates an inner product in spin space.
 This means that the scattering matrix is diagonal in
 angular-momentum basis:
 \beq
 M^{(NR)}_{JMl;J^{\prime}M^{\prime}l^{\prime}}&=&
 M^{(NR)}_{J^{\prime}M^{\prime}l^{\prime};JMl}
\nonumber\\
 &=&\delta_{J^{\prime}J}\delta_{M^{\prime}M}\delta_{l^{\prime}l}\frac{1}{2ik}(S^{Jl}-1)
 \;.
 \label{define_M_nr}
 \eeq
 For single channel spin-dependent scattering, the $S$-matrix is parameterized
 by $S^{Jl}=e^{2i\delta_{Jl}}$.
 Since $J$ can take two possible values, $J=l\pm 1/2$, we will also
 conveniently denote them as
 $S^{J=l\pm1/2,l}=S^{l\pm}=\exp(2i\delta_{l\pm})$.

 If we have chosen the $z$-axis to coincide with the
 incident momentum $\mathbf{\hat{k}}$,
 the scattering amplitude depends only on $\theta$:
  \bw
 \beq
 M^{(NR)}_{s^{\prime}s}(\theta)
 &=&\frac{1}{2ik}\sum_{l=0}^{\infty}\sum_{J=l-\frac{1}{2}}^{l+\frac{1}{2}}\sqrt{4\pi(2l+1)}(S^{Jl}-1)
 S^{l\frac{1}{2}}_{JM;ms^{\prime}}
 S^{l\frac{1}{2}}_{JM;0s}
 Y_{lm}(\mathbf{\hat{r}})\;.\label{scattering_amplitude_2}
 \eeq
 \ew
 Due to the Clebsch-Gordan coefficients
 $S^{l\frac{1}{2}}_{JM;0s}$ and $S^{l\frac{1}{2}}_{JM;ms^{\prime}}$, we find:
 $M=s$ and $m=M-s^{\prime}$.
 Thus, in the previous equation we ignore the sum over  $M$ and $m$
 for a given pair of $s$ and $s^{\prime}$.
 It is then clear that $M(\theta)$ as a matrix in spin space
 can also be written in the following form
 \be
 M(\theta)=f(\theta)+ig(\theta)(\bsigma\cdot\mathbf{e})\;,
 \ee
 where $f(\theta)$ and $g(\theta)$ are known as
 no-flip and spin-flip amplitudes, respectively.
 $\bsigma$ is the Pauli matrices in spin space
 and $\mathbf{e}$ is a unit vector perpendicular to the scattering plane.
 With this convention, these functions are given by
 \begin{equation}
 \left\{ \begin{aligned}
 f(\theta) &=  \sum_{l=0}^\infty
 [(l+1)f_{l+}
 +lf_{l-}]P_l(\cos\theta)\\
 g(\theta) &= \sum_{l=0}^\infty [f_{l+}-f_{l-}]P^1_l(\cos\theta)
 \end{aligned} \right.\;.
 \end{equation}
 with $f_{l\pm}(k)=e^{i\delta_{l\pm}}\sin(\delta_{l\pm})/k$.
 These are identical to what have appeared in the early literature,
 see for example, Ref.~\cite{Hamilton:1963} and
 Ref.~\cite{chew:1957}.

 Let us now enclose the system we considered in a large
 cubic box with periodic boundary conditions applied in all
 three spatial directions. The interaction is only present
 in the inner region while in the outer region, the radial wave function
 becomes a superposition of $n_l(kr)$ and $j_l(kr)$,
 \be
 \psi(\mathbf{r})
 =\sum_{JMl}(\alpha_{Jl}(k)j_{l}(kr)
 +\beta_{Jl}(k)n_{l}(kr))Y_{JM}^{l\frac{1}{2}}(\mathbf{\hat{r}})
 \label{wavefunction3}
 \ee
 which leads to the identification
 $\tan(\delta_{Jl}(k))=\beta_{Jl}(k)/\alpha_{Jl}(k)$.
 The Green's functions, which are singular periodic solutions of the
 Helmholtz equation, has a similar expansion. In this case, it is similar to those introduced in
 Ref.~\cite{luscher91:finitea} (see also e.g.
 Ref.~\cite{Bernard:2008ax,Ishizuka:2009bx}):
  \bw
  \be
 G_{JMl}(\textbf{r},k^2)=
 \frac{(-1)^lk^{l+1}}{4\pi}\left(Y_{JM}^{l\frac{1}{2}}(\mathbf{{\hat{r}}})n_l(kr)+\underset{J^{\prime}M^{\prime}l^{\prime}}{\sum}
 \mathcal{M}_{JMl;J^{\prime}M^{\prime}l^{\prime}}(k^2)
 Y_{J^{\prime}M^{\prime}}^{l^{\prime}\frac{1}{2}}
 (\mathbf{{\hat{r}}})j_{l^{\prime}}(kr)
 \right)\;.
 \label{wavefunction4}
 \ee
 \ew
 Comparing Eq.~(\ref{wavefunction3}) with Eq.~(\ref{wavefunction4})
 and following similar steps as in the derivation of the conventional L\"uscher's
 formula, one finally finds
 \be
 \det[\tan\delta_{Jl}(k)\mathcal{M}_{JMl;J^{\prime}M^{\prime}l^{\prime}}
 -\delta_{JJ^{\prime}}\delta_{MM^{\prime}}\delta_{ll^{\prime}}]=0
 \ee
 which is the same as in Ref.~\cite{Bernard:2008ax}.
 The explicit form of $\mathcal{M}_{JMl;J^{\prime}M^{\prime}l^{\prime}}$ is given in Ref.~\cite{Bernard:2008ax}
 which we quote here:
 \bw
 \beq
 \label{eq:M_JLM}
 \mathcal{M}_{JMl;J^{\prime}M^{\prime}l^{\prime}}&=&
 \sum_{mm^{\prime}s}\mathcal{M}_{lm;l^{\prime}m^{\prime}}\langle{lm\frac{1}{2}s}|JM\rangle
 \langle{l^{\prime}m^{\prime}\frac{1}{2}s}|J^{\prime}M^{\prime}\rangle \;,
 \eeq
 \beq
 \mathcal{M}_{lm;l^{\prime}m^{\prime}} &=&
 \frac{(-1)^{l}}{\pi^{\frac{3}{2}}}\sum_{t=|l-l^{\prime}|}^{l+l^{\prime}}
 \sum^{t}_{n=-t}\frac{i^t}{\boldsymbol{\kappa}^{t+1}}Z_{tn}(1;\boldsymbol{\kappa}^2)
 C_{lm,tn,l^{\prime}m^{\prime}}
 \eeq
 \ew
 with $\boldsymbol{\kappa}=\mathbf{k}L/(2\pi)$ and the zeta function $Z_{tn}(1;\boldsymbol{\kappa}^2)$
 and the coefficients $C_{lm,tn,l^{\prime}m^{\prime}}$
 are given respectively by
 \bw
 \beq
 \label{eq:Z_jn}
 Z_{tn}(1;\boldsymbol{\kappa}^2)&=&\sum_{\mathbf{n}\in{\mathbbm{Z}^{3}}}\frac{|\mathbf{n}|^t
 Y_{tn}(\mathbf{\hat{n}})}{\mathbf{n}^2-\boldsymbol{\kappa}^2}\;,
 \eeq
 \beq
 C_{lm,tn,l^{\prime}m^{\prime}}&=&i^{l-t+l^{\prime}}\sqrt{\frac{(2l+1)(2t+1)}{(2l^{\prime}+1)}}
 \langle{l0t0}|l^{\prime}0\rangle
 \langle{lmtn}|l^{\prime}m^{\prime}\rangle\;.
 \eeq
 \ew

 The corresponding formulae in moving frames have been obtained in
 Ref.~\cite{Rummukainen:1995vs,ZiwenFu2012}. For example, instead of
 Eq.~(\ref{eq:M_JLM}), we have,
 \be
 \label{eq:Md0}
 \mathcal{M}^{\mathbf{d}}_{lm;l^{\prime}m^{\prime}}=
 \frac{(-1)^{l}}{\gamma\pi^{\frac{3}{2}}}\sum_{t,n}
 \frac{i^t}{\boldsymbol{\kappa}^{t+1}}
 Z^{\mathbf{d}}_{tn}(1;\boldsymbol{\kappa}^2)
 {C_{lm,tn,l^{\prime}m^{\prime}}}
 \ee
 with $\boldsymbol{\kappa}=\mathbf{k}L/(2\pi)$, the parameter $\gamma$ is the
 Lorentz boost factor associated with the moving-frame
 and the zeta function $Z^{\mathbf{d}}_{tn}(1;\boldsymbol{\kappa}^2)$ is defined by
 \beq
 Z^{\mathbf{d}}_{tn}(1;\boldsymbol{\kappa}^2)&=&
 \sum_{\mathbf{n}\in\mathcal{P}_\mathbf{d}}\frac{|\mathbf{n}|^tY_{tn}(\mathbf{\hat{n}})}{\mathbf{n}^2-\boldsymbol{\kappa}^2}\;, \label{zeta}
 \eeq
 the coefficients $C_{lm,tn,l^{\prime}m^{\prime}}$ are the same as
 in non-moving frames, e.g. Eq.~(\ref{eq:Z_jn}).
 In the above formulae,
 \be
 \mathcal{P}_{\mathbf{d}}=\left\{\mathbf{r}\left|\mathbf{r}
 =\vec{\gamma}^{-1}(\mathbf{n}+\frac{1}{2}\alpha{\mathbf{d}})
 \;\;\; \mathbf{n}\in{\mathbbm{Z}^{3}}\right. \right\}\;,
 \ee
 where $\alpha$ and $\mathbf{d}$ are given as in
 subsection~\ref{subsec:MF}.
 Then Eq.~(\ref{eq:M_JLM}) still holds except that one has to use
 the moving-frame versions ($\mathcal{M}^{\mathbf{d}}_{JMl;J^{\prime}M^{\prime}l^{\prime}}$
 and $\mathcal{M}^{\mathbf{d}}_{lm;l^{\prime}m^{\prime}}$) in place of the non-moving frame
 versions.

 Extension of the above formulae to the case of multi-channel case is straightforward.
 For example, Eq.~(\ref{define_M_nr}) is generalized to
 \beq
 M^{(NR)}_{ij;JMl;J^{\prime}M^{\prime}l^{\prime}}&=&
 M^{(NR)}_{ij;J^{\prime}M^{\prime}l^{\prime};JMl}\nonumber\\
 &=&\delta_{J^{\prime}J}\delta_{M^{\prime}M}
 \delta_{l^{\prime}l}\frac{1}{2i\sqrt{k_ik_j}}(S_{ij}^{Jl}-\delta_{ij})\;.\nonumber\\          \label{define_multi_M_nr}
 \eeq

 Similarly, Eq.~(\ref{eq:M_JLM}) is modified to:
 \be
 \label{eq:Md1}
 \mathcal{M}^{\mathbf{d}}_{i;JMl;J^{\prime}M^{\prime}l^{\prime}}=
 \sum_{mm^{\prime}s}\mathcal{M}^{\mathbf{d}}_{i;lm;l^{\prime}m^{\prime}}\langle{lm\frac{1}{2}s}|JM\rangle
 \langle{l^{\prime}m^{\prime}\frac{1}{2}s}|J^{\prime}M^{\prime}\rangle
 \ee
 with $\mathcal{M}^{\mathbf{d}}_{i;lm;l^{\prime}m^{\prime}}$ given in terms
 of the zeta function
 $Z^{\mathbf{d}}_{i;tn}(1;\boldsymbol{\kappa_i}^2)$ like
 Eq.~(\ref{eq:Md0}),
with $\boldsymbol{\kappa}_i=\mathbf{k_i}L/(2\pi)$.

\section{Calculation the the kinematic factor of loop integration/summation in a single channel}
 \label{appendix:loops}
 In this appendix,
 we use the notation: $k^{\ast}=|\mathbf{k^{\ast}}|$, $q^{\ast}=|\mathbf{q^{\ast}}|$
 unless otherwise stated.

 The generic finite-volume corrections that we are interested in
 has the following form:
 \beq
 S(\mathbf{q^{\ast}})&=&\frac{1}{L^3}\sum_{\mathbf{k}}\frac{w_\bk^{\ast}}{w_\bk}\frac{f^{\ast}(\mathbf{k}^{\ast})}{q^{{\ast}2}-k^{{\ast}2}}\;,
 \eeq
 where the function $f^{\ast}(\mathbf{k}^{\ast})$
 has no singularities for real $\mathbf{k}^{\ast}$ and
 falls off fast enough at $k^{\ast}\rightarrow \infty$ so as to
 render the summation convergent. To simplify the matter, we assume
 that it is spin-independent and thus can be expanded into spherical harmonics
 as in Ref.~\cite{Kim:2005gf}:
  \be
 f^{\ast}(\mathbf{k}^{\ast})=\sum_{l=0}^{\infty}\sum_{m=-l}^{l}f^{\ast}_{lm}(k^{\ast})k^{{\ast}l}\sqrt{4\pi}Y_{lm}(\mathbf{\hat{k}}^{\ast})\;.
 \label{definition_testfunction}
 \ee
 One would like to study the behavior
 of $S(\mathbf{q^{\ast}})$ in the limit of large $L$.
 For a fixed $L$, if there is no term in the sum with $k^{{\ast}2}\simeq q^{{\ast}2}$,
 we can replace the sum by an integration. The singularity
 at $k^{{\ast}2}\simeq q^{{\ast}2}$ for large $L$ forbids this simple replacement.
 Basically $S(\mathbf{q^{\ast}})$ will split into two parts,
 one of which can be approximated by the principle-valued integral,
 the other being the finite volume correction.
 This has been established in Ref.~\cite{Kim:2005gf}
 and we directly quote their final results:
 \beq
 S(\mathbf{q^{\ast}})&=&\mathcal{P}\int\frac{d^3k^{\ast}}{(2\pi)^3}\frac{f^{\ast}(\mathbf{k}^{\ast})}{q^{{\ast}2}-k^{{\ast}2}}
 +\sum_{l,m}f_{lm}C^P_{lm}(q^{*2})\;,
  \label{testfunction}\nonumber\\
  \eeq
  \beq
C^P_{lm}(q^{{\ast}2})&=&\frac{1}{L^3}\sum_{\mathbf{k}}\frac{w_{\mathbf{k}}^{\ast}}{w_{\mathbf{k}}}
\frac{e^{\alpha(q^{{\ast}2}-k^{{\ast}2})}}{q^{{\ast}2}-k^{{\ast}2}}k^{{\ast}l}\sqrt{4\pi}Y_{lm}(\mathbf{\hat{k}}^{\ast})\nonumber\\
& &-\mathcal{P}\int\frac{d^3k^{\ast}}{(2\pi)^3}\frac{e^{\alpha(q^{{\ast}2}-k^{{\ast}2})}}{q^{{\ast}2}-k^{{\ast}2}}\;,
\eeq
 where $\mathcal{P}$ stands for the principal-value prescription.
 This summation formula will be utilized shortly.

 \begin{figure}[h]
 \centering
 \includegraphics[width=0.5\textwidth]{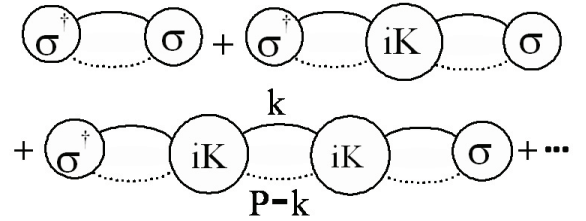}\caption{The dash line represents the meson propagator, and
 the solid line represents the baryon propagator.
 A circle with a symbol $iK$ inside stands for the Bethe-Salpeter kernel
 which consists of all amputated two-particle irreducible diagrams.
 A circle with a symbol $\sigma$ or $\sigma^\dagger$ denotes
 the interpolating operators in Eq.~\ref{eq:CP_definition}.}
 \label{fig:fig2}
 \end{figure}
 We now come to the correlation function $C(P)$ defined
 in Eq(\ref{eq:CP_definition}). It can be expressed in terms of
 the Bethe-Salpeter kernel $iK$ through the series shown
 in Fig.~\ref{fig:fig2}.
 The loop integration/summation appearing in the figure has the
 following form:
 \be
 I=\frac{-1}{L^{3}}\sum_{\mathbf{{k}}}\int \frac{dk_{0}}{2\pi}\frac{f(k_{0},\mathbf
 {k})(k^{\mu}\gamma_{\mu}+m_{1})}{(k^{2}-m_{1}^{2}+i\varepsilon)
 ((P-k)^{2}-m_{2}^{2}+i\varepsilon)}\;,
 \ee
 where $k=(k_0,\mathbf{k})$, $P=(E,\mathbf{P})$ being the
 corresponding four-momenta and
 $m_{1}$ and $m_{2}$ being the masses of the baryon and the meson
 respectively.
 The function $f(k)$ contains the energy-momentum dependence arising from the kernels as well
 as that from the dressed propagators. The properties of $f(k)$ is such that there
 exists no singularities for real $\mathbf{k}$ and its ultraviolet
 behavior is to render the integration or summation convergent.
 Integrating out $k_{0}$ one gets
 \bw
 \be
 I =\frac{-i}{L^{3}}\sum_{\mathbf{k}}\left(
 \frac{f(w_{1\mathbf{k}},\mathbf{k})C(\mathbf{k})}
 {2w_{1\mathbf{k}}((E-w_{1\mathbf{k}})^{2}-w_{2\mathbf{k}}^{2})}
 +\frac{f(E+w_{2\mathbf{k}},\mathbf{k})D(\mathbf{k})}{2w_{2\mathbf{k}}
 ((E+w_{2\mathbf{k}})^{2}-w_{1\mathbf{k}}^{2})}\right)\;.
 \label{loop_sum_1}
 \ee
 \ew
 If we assume $f(\bk)$ is an even function for $\bk$
 \be
  \label{eq:def_of_CD}
 \left\{ \begin{aligned}
 C(\mathbf{k}) &=(k^{0}\gamma_{0}+m_{1})|_{k_{0}=w_{1\mathbf{k}}}
 \\
 D(\mathbf{k})&=(k^{0}\gamma_{0}+m_{1})|_{k_{0}=E+w_{2\mathbf{k}}}
 \end{aligned} \right.\;,
 \ee
 and $w_{1\mathbf{k}}=\sqrt{\mathbf{k}^{2}+m_{1}^{2}}$,
 $w_{2\mathbf{k}}=\sqrt{(\mathbf{P}-\mathbf{k})^{2}+m_{2}^{2}}$ are the
 two energies. Note that both $C(\mathbf{k})$ and $D(\mathbf{k})$ and
 thus the integral $I$ are matrices in Dirac space.
 However, as mentioned in the main text, normally the interpolating
 operators that we formed to create the meson and baryon carry
 definite parity. This means that, for practical applications,
 $\gamma_0$ takes its eigenvalues, $\pm 1$ times a unit matrix (in spin space).
 Thus the integral $I$ is written into two parts $I_{1}$ and
 $I_{2}$ corresponding to the two terms in Eq(\ref{loop_sum_1}).
 The second term $I_2$ does not contain the finite-volume
 singularities in the kinematic region of interest and therefore
 can be replaced by the corresponding integral in the large volume
 limit. The term which does contain the two-particle finite-volume
 poles is $I_1$,
 \be
 I_{1}=\frac{-i}{L^{3}}\sum_{\mathbf{k}}\frac{f(w_{1\mathbf{k}},\mathbf{k})C(\mathbf{k})}
 {2w_{1\mathbf{k}}((E-w_{1\mathbf{k}})^{2}-w_{2\mathbf{k}}^{2})}
 \;.
 \ee
 It is seen that the two-particle pole singularity in $I_1$
 is located at $E=w_{1\mathbf{k}}+w_{2\mathbf{k}}$.
 To determine the finite volume correction in more detail,
 we express the term $I_{1}$ in another form by transforming it in to the COM frame.
 In COM frame the two energies are:
 $w^{\ast}_{1\mathbf{k}}=\sqrt{\mathbf{k}^{{\ast}2}+m_{1}^{2}}$,
 $w^{\ast}_{2\mathbf{k}}=\sqrt{\mathbf{k}^{{\ast}2}+m_{2}^{2}}$
 as given in section~\ref{subsec:MF}.Thus we obtain:
 \begin{align}
 I_{1} &  =-\frac{i}{L^{3}}\sum_{\mathbf{k}}\frac{f^{\ast}(\mathbf{k}^{\ast})
 C(\mathbf{k}^{\ast})}{2w_{1\mathbf{k}}((E^{\ast}-w_{1\mathbf{k}}^{\ast})^{2}-w_{2\mathbf{k}}^{\ast2})}\nonumber\\
 &  =\frac{-i}{2L^{3}}\sum_{\mathbf{k}}\frac{w_{1\mathbf{k}}^{\ast}}{w_{1\mathbf{k}}}
 \frac{f^{\ast}(\mathbf{k}^{\ast})C(\mathbf{k}^{\ast})}{q^{\ast2}-k^{\ast2}}
 \left[\frac{(E^{\ast}+w_{1\mathbf{k}}^{\ast})^{2}-w_{2\mathbf{k}}^{\ast2}}
 {4E^{\ast 2}w_{1\mathbf{k}}^{\ast}
 }\right]\;.
 \label{loop_integer_1}
 \end{align}

 By using the summation formula Eq.~(\ref{testfunction}) we mentioned
 at the beginning of this appendix, we obtain:
 \begin{widetext}
 \beq
 \label{eq:I_1_final}
 I_{1}&=& (-i)\mathcal{P}\int \frac{d^{3}k^{\ast}}{(2\pi)^{3}}
 \frac{f^{\ast}(\mathbf{k}^{\ast})C(\mathbf{k}^{\ast})}{q^{\ast2}-k^{\ast2}}
 \left[\frac{(E^{\ast}+w_{1\mathbf{k}}^{\ast})^{2}-w_{2\mathbf{k}}^{\ast2}}
 {8E^{\ast 2}w_{1\mathbf{k}}^{\ast}}\right]
 -\frac{iC(\mathbf{q^{\ast}})}{2E^{\ast}}\sum_{lm}f_{lm}^{\ast}(q^{\ast})C_{lm}^{P}(q^{\ast2})\;.
 \eeq
 \end{widetext}
 In order to write $I_{1}$ as the infinite-volume result together
 with a correction, we replace the principle-value integration
 in the above formula by a Feynman $+i\varepsilon$ prescription in
 the propagator and a ``delta-function" term which picks out the $l=0$
 part of $f^{\ast}$.
 \begin{widetext}
 \begin{align}
 I_{1}
 &  =-i\frac{1}{2E^{\ast}}\int \frac{d^{3}k^{\ast}}{(2\pi)^{3}}
 \frac{f^{\ast}(\mathbf{k}^{\ast})C(\mathbf{k}^{\ast})}{q^{\ast2}-k^{\ast2}+i\varepsilon}
 \frac{(E^{\ast}+w_{1\mathbf{k}}^{\ast})^{2}-w_{2\mathbf{k}}^{\ast2}}{4E^{\ast}w_{1\mathbf{k}}^{\ast}}
 +\frac{q^{\ast}f_{00}^{\ast}(q^{\ast})C(\mathbf{q^{\ast}})}{8\pi{E^{\ast}}}
 -\frac{iC(\mathbf{q^{\ast}})}{2E^{\ast}}\sum_{lm}f_{lm}^{\ast}(q^{\ast})C_{lm}^{P}(q^{\ast2})\;.
 \end{align}
 \end{widetext}
 Finally, we arrive at our final result for $I$,
 \beq
 I&=&I_{\infty}+I_{FV} \label{loop_2}\;,
 \eeq
 where $I_{\infty}$ is the infinite volume result for the original loop
 integral with the appropriate Feynman's prescription,
 $I_{\infty}=I_{1,\infty}+I_2$.
 As mentioned earlier, $I_{\infty}$  contains no finite volume singularities.
 The two-particle singularities are contained in
 the finite volume correction term, $I_{FV}$ which is given by
 \beq
 I_{FV}&=&\frac{q^{\ast}f_{00}^{\ast}(q^{\ast})C(\mathbf{q^{\ast}})}{8\pi{E^{\ast}}}
 -\frac{iC(\mathbf{q^{\ast}})}{2E^{\ast}}\sum_{lm}f_{lm}^{\ast}(q^{\ast})C_{lm}^{P}(q^{\ast2})\;.\nonumber\\    \label{I_fv}
 \eeq

 Using the expansion~(\ref{definition_testfunction}) and
 the completeness of spherical harmonics, we obtain,
 \bw
 \beq
 \label{field_F3}
 F&\equiv &C(\mathbf{q^{\ast}})\tilde{F}
 =\frac{q^{\ast}C(\mathbf{q^{\ast}})}{8\pi{E^{\ast}}}(1+iF^{FV})\;,
 \eeq
 \beq
 \tilde{F}_{JMl;J^{\prime}M^{\prime}l^{\prime}}&=&\frac{q^{\ast}}{8\pi
 {E^{\ast}}}(\delta_{JJ^{\prime}}\delta_{MM^{\prime}}\delta_{ll^{\prime}}
 +iF_{JMl;J^{\prime}M^{\prime}l^{\prime}}^{FV})\;,    \label{field_F2}
 \eeq
 \beq
 F_{JMl;J^{\prime}M^{\prime}l^{\prime}}^{FV}&=&\frac{-4\pi}{q^{\ast}}
 \sum_{l_1m_1}\frac{\sqrt{4\pi}}{q^{\ast l_{1}}}{C_{l_1m_1}^{P}}\int{d\Omega^{\ast}}Y_{JM}^{l\frac{1}{2}\dagger}
 Y_{l_1m_1}^{\ast}Y_{J^{\prime}M^{\prime}}^{l^{\prime}\frac{1}{2}}\;, \label{field_F1}
 \eeq
 \beq
 C^P_{l_1m_1}(q^{{\ast}2})&=&\frac{1}{L^3}\sum_{\mathbf{k}}\frac{w_{1\mathbf{k}}^{\ast}}{w_{1\mathbf{k}}}
 \frac{e^{\alpha(q^{{\ast}2}-k^{{\ast}2})}}{q^{{\ast}2}-k^{{\ast}2}}k^{{\ast}l_1}\sqrt{4\pi}Y_{l_1m_1}
 -\mathcal{P}\int\frac{d^3k^{\ast}}{(2\pi)^3}\frac{e^{\alpha(q^{{\ast}2}-k^{{\ast}2})}}{q^{{\ast}2}-k^{{\ast}2}}\;,
 \eeq
 \ew
 where $\mathcal{P}$ stands for the principal-value prescription.

 In Ref.~\cite{Briceno:2012yi}, for the scattering of two particles with different masses,
 a relation between $C^P_{lm}(q^{{\ast}2})$ and $Z^\mathbf{d}_{lm}(1;\boldsymbol{\kappa}^2)$ is found,
 \beq
 C^{P}_{lm}(q^{{\ast}2})=-\frac{\sqrt{4\pi}}{\gamma{L^3}}(\frac{2\pi}{L})^{l-2}Z^\mathbf{d}_{lm}(1;\boldsymbol{\kappa}^2)
 \eeq
 with $\boldsymbol{\kappa}=\mathbf{q^{\ast}}L/(2\pi)$, $Z^{\mathbf{d}}_{lm}(1;\boldsymbol{\kappa}^2)$ is defined by (\ref{zeta}).
 This equality holds up to terms which vanish exponentially with the box size.
 Thus, according to the expressions for $F^{FV}_{JMl;J^{\prime}M^{\prime}l^{\prime}}$
 and $\mathcal{M}^{\mathbf{d}}_{JMl;J^{\prime}M^{\prime}l^{\prime}}$ i.e. Eq.~(\ref{field_F1}) and Eq.~(\ref{eq:Md1})
 we have the following equality:
 \beq
 \label{equality_1}
 F^{FV}_{JMl;J^{\prime}M^{\prime}l^{\prime}}&=&i^{l-l^{\prime}}\mathcal{M}^{\mathbf{d}}_{JMl;J^{\prime}M^{\prime}l^{\prime}}
 \eeq
 which is also valid up to terms that vanish exponentially with the box size.
 In the case of multi-channel scattering,
 the above formulae is naturally modified to:
 \beq
 C^{P}_{i;lm}(q_i^{\ast2})=-\frac{\sqrt{4\pi}}{\gamma{L^3}}(\frac{2\pi}{L})^{l-2}Z^{\mathbf{d}}_{i;lm}(1;\boldsymbol{\kappa}_i^2)
 \eeq
 with $\boldsymbol{\kappa}_i=\mathbf{q_i^{\ast}}L/(2\pi)$
 and Eq.~(\ref{equality_1}) is modified to
\beq
 \label{equality_2}
 F^{FV}_{i;JMl;J^{\prime}M^{\prime}l^{\prime}}&=&i^{l-l^{\prime}}\mathcal{M}^{\mathbf{d}}_{i;JMl;J^{\prime}M^{\prime}l^{\prime}}\;.
 \eeq
 Again, these formulae hold up to terms that are vanishing
 exponentially in the box size. These equalities are utilized
 when we compare L\"uscher's formulae obtained from quantum field
 theory with those obtained from non-relativistic quantum mechanics.

%

\end{document}